\documentclass[12pt]{iopart}

\usepackage{graphicx}
\usepackage{amsthm}
\usepackage{amssymb}

\usepackage{color}
\usepackage{epstopdf}
\usepackage{latexsym}
\usepackage{bm}

\usepackage{subfigure}

\newcommand{\ket}[1]{|{#1}\rangle}

\newcommand{\bra}[1]{\langle{#1}|}

\newcommand{\kets}[2]{|{#1}\rangle_{#2}\hspace*{-0.2mm}}

\newcommand{\ketbras}[3]{\ket{#1}_{#3}\hspace*{-0.mm}\bra{#2}}

\newcommand{\bracket}[2]{\langle#1|#2\rangle}
\renewcommand{\Re}{\mathop{\mathrm{Re}}\limits}

\newcommand{\Cov}{\mathop{\mathrm{Cov}}\limits}
\newcommand{\Var}{\mathop{\mathrm{Var}}\limits}
\newcommand{\diag}{\mathop{\mathrm{diag}}\limits}

\newcommand{\s}{\sigma}
\definecolor{dgreen}{rgb}{0,0.5,0}

\newcommand{\blue}{\color{blue}}

\definecolor{delete}{cmyk}{0.5,0,0,0}

\begin{document}

\title{Entanglement-assisted tomography of a
quantum target}


\author{A. De Pasquale$^{1}$, P. Facchi$^{2,3}$, V. Giovannetti$^{1}$, K. Yuasa$^{4}$}
\address{$^1$NEST, Scuola Normale Superiore and Istituto Nanoscienze-CNR, 
I-56126 Pisa, Italy}     
\address{$^2$Dipartimento di Matematica and MECENAS, Universit\`a di Bari, I-70125 Bari, Italy}
\address{$^3$INFN, Sezione di Bari, I-70126 Bari, Italy}
\address{$^4$Waseda Institute for Advanced Study, Waseda University, Tokyo 169-8050, Japan}



\date \today

\begin{abstract}
We study the efficiency of quantum tomographic reconstruction where 
the system under investigation (quantum target) is indirectly monitored  
by looking at  the state of a quantum probe  that  has been scattered off the target. 
In particular we focus on the state tomography of a qubit through a one-dimensional scattering of a probe qubit, with a Heisenberg-type interaction.
Via direct evaluation of the associated quantum Cram\'{e}r-Rao bounds, 
we compare the accuracy efficiency that one can get by adopting 
entanglement-assisted  strategies with that achievable
when entanglement resources are not available. 
Even though  sub-shot noise accuracy levels are not attainable, we show that    
quantum correlations play a significant role in the  estimation. A comparison with
the accuracy levels obtainable by direct estimation (not through a probe) of the quantum target is also performed.  
\end{abstract}


\maketitle
\section{Introduction}
 The possibility of reconstructing the quantum state of a system via measurements 
(quantum state tomography, QST) is a central problem in quantum
information theory~\cite{PARIS}, which poses a series of fundamental questions related
to the fact that the state itself is not directly observable and that each given quantum measurement
typically reveals only partial information on the observed system.     
In recent years, a great deal of work has been 
devoted to this issue and many important features have been 
recognized, including the fact  that having at disposal several copies (say $M$)   of the initial state, collective measurements are more informative than individual ones,
e.g., see Refs.~\cite{PARIS,VARI,bagan} and references therein. 
In abstract terms, QST ultimately reduces to the ability of estimating the set  of continuos  parameters 
which  define the expansion of an unknown state  with respect to a 
reference basis of operators (say the set of Pauli matrices for a two-level system, qubit). As such, its ultimate
accuracy limits can be evaluated by exploiting some general results of quantum estimation theory~\cite{HELS67,HOLE11,ParisRev,ADVquantmetr} 
(more precisely, of a part of the theory which directly  deals with the estimation of 
continuos parameters).

Up to date, most of the works focused on scenarios where the system under investigation
can be directly accessed, i.e., posing no constraint whatsoever to the physical operations one may
perform on it. In this context, for instance, the ultimately accuracy limit obtainable in the tomographic
reconstruction of a qubit initialized in an arbitrary (possibly) mixed state 
have been set  in Ref.~\cite{bagan}, by computing the associated quantum Cram\'{e}r-Rao (CR) bound~\cite{HELS67,HOLE11},  in terms of the quantum Fisher information (QFI) matrix  of the problem 
(see below for the precise definitions). 
In this paper, instead, we address the problem from a slightly different perspective, which 
captures an important aspect of many realistic experimental situations. 
More precisely,  along the line
set in Ref.~\cite{paper:scattering},  we 
consider the case in which the system of interest (from now on the \emph{target} $X$),
can only be addressed indirectly via measurements performed on 
a probe which has interacted with it.  In our model the latter is described as a quantum system  $A$
 characterized both by external (e.g., momentum/position) and internal (e.g., spin) degrees of freedom, which 
the experimentalists are allowed to prepare in any initial configuration (also the target 
possesses external {\em and} internal degrees of freedoms but, for the sake of simplicity, only these
last are supposed to be unknown, the external degrees of freedom being assigned by fixing the
position of the target system). 
The tomographic reconstruction  then proceeds by letting 
 $A$ and $X$ interact via a scattering process and by measuring the 
final state of the former (or at least a part of  it which has been scattered along some
preferred direction). 
Indeed, the whole setting is devised 
in order to mimic the basic features of a standard (Rutherford-like) scattering experiment, where
one tries to reconstruct the properties of a target system by firing  probe particles on it
and by looking at the way they emerge from the process. For the sake of simplicity
we will limit the analysis to  the case of 1D scattering processes, and describe the internal
degrees of freedom of $A$ and $X$ as two-level (spin) systems (similar models have been recently analyzed to study entanglement generation~\cite{quantControl,entanglementGeneration}).

\begin{figure}
\begin{center}
\includegraphics[height=0.27\columnwidth]{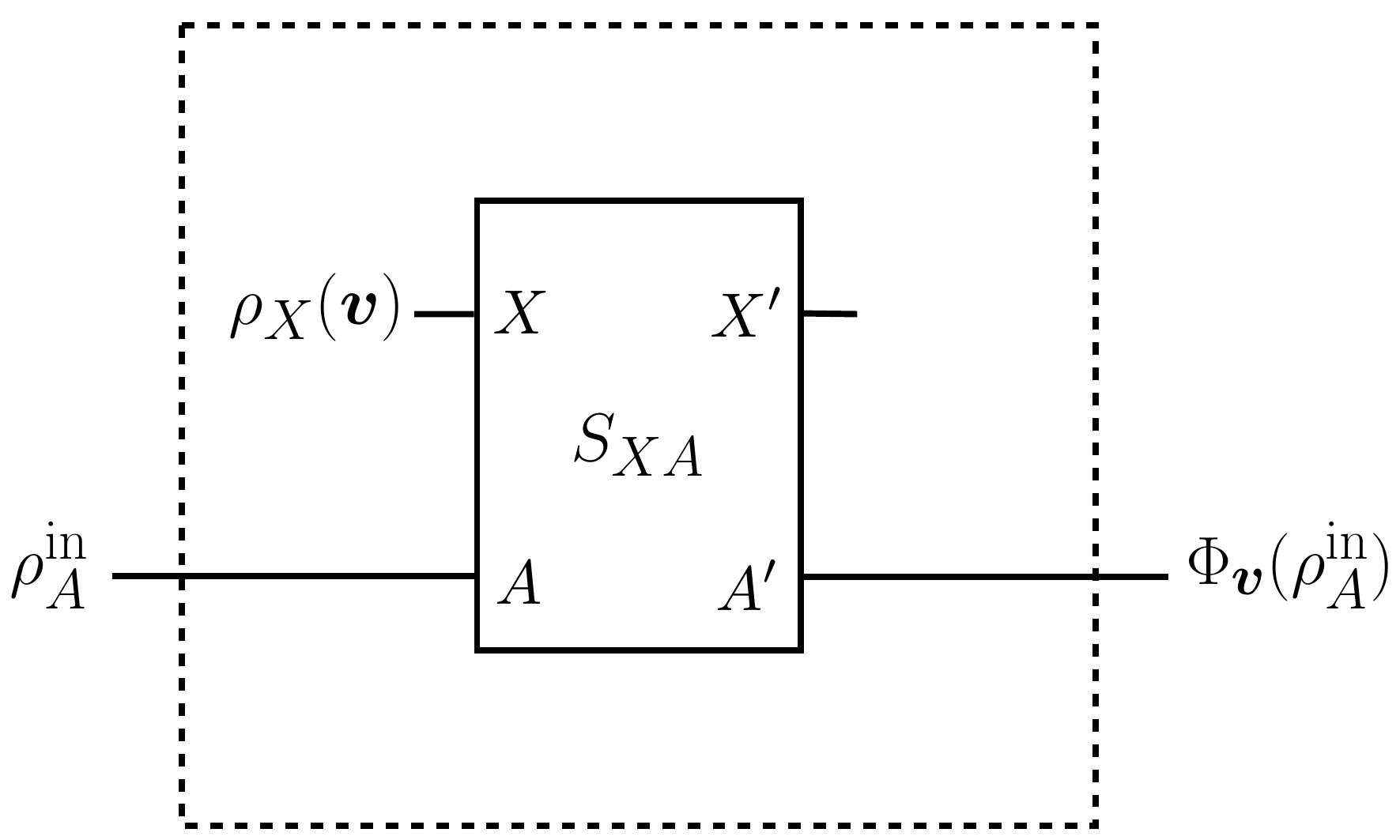}
\caption{Scheme of a programmable quantum channel.
Here the CPTL map $\Phi_{\bm{v}}$ is
defined as an interaction with an external system $X$ through a given (fixed) coupling represented by the operator~$S_{XA}$. 
In our problem
the channel to be reconstructed describes the 1D scattering of a probe $A$ off a target qubit $X$, with Heisenberg-type interaction.
This process induces a modification of the probe initial state
$\rho_A^{\mathrm{in}}$ according to the target initial state $\rho_X(\bm{v})$
identified  by the coordinates ${\bm{v}}$.}
\label{fig:programmableChannel}
\end{center}
\end{figure}

 It is worth noticing that  the problem we are considering admits also
an interpretation in the context of \emph{quantum channel estimation} theory (see Ref.~\cite{ADVquantmetr} and references therein). 
This is the theory, sometimes
identified with the  name of \emph{quantum metrology},  which studies the efficiency of those schemes designed to recover information not on a quantum system, but on a quantum channel (quantum process tomography, QPT).
We remind that in quantum mechanics quantum channels represent the most general physical transformations and are   fully described by assigning completely-positive trace-preserving linear (CPTL) mappings~\cite{HOLE11}, which act on the density matrices of the system (in our case 
the probe $A$).  
In quantum metrology   the 
mapping $\Phi_{\bm{v}}$ is assumed to  belong to a family of transformations identified  by 
a set of parameters  $\bm{v}$, whose values are unknown and which we wish to recover by
preparing the system in some fiduciary initial state $\rho_A^{\mathrm{in}}$ and
by measuring the corresponding output state transformed by the channel. 
 In our case the quantum channel to be estimated is the one that 
induces a modification on the probe $A$ via its interaction with the target $X$, while the
$\bm{v}$'s correspond  to the 
parameters that define the (unknown) state $\rho_X$  
(see Fig.~\ref{fig:programmableChannel}). 
In the jargon introduced in 
Ref.~\cite{progChannels}, this transformation belongs to the special class of {\em programmable 
channels}\footnote{Explicitly, these channels can indeed be parametrized by assigning a fixed interaction with an external unknown system. In a seminal paper \cite{NielsenChuangProgr} Nielsen and Chuang proved that the family of all unitaries acting on $n$ qubits is not programmable, since an $n$-qubit register can encode at most $2^n$ distinct quantum operations. Notwithstanding the non-universality of programmable channels, the authors also pointed out the possibility to program the family of unitary operations probabilistically. In this context, an interesting model was proposed for the case of single qubits in \cite{Vidal} and implemented for photonic qubits in \cite{Micuda}.}.
A well-known fact is that, in general,  if $M$ is the number of tests we perform in order
to recover the actual values of $\bm{v}$
 (each test consisting of applying the \emph{same} channel $\Phi_{\bm{v}}$ to 
a new copy of probe $A$)
 the statistical  scaling of the associated uncertainty  can be reduced from the ``standard quantum limit'' (SQL) (or ``shot noise'' in quantum optics) $1/\sqrt{M}$ scaling, to the so-called ``Heisenberg bound'' $1/M$ scaling, by the introduction of suitable quantum correlations between 
 initializations of the various copies of $A$~\cite{ADVquantmetr,quantmetr}.
   However, this is not the case if, as in our case,  the channels under investigation  are programmable~\cite{progChannels}. 
   As a consequence in our model no sub-shot noise scaling in $M$ of  the accuracy should be expected. For this reason we will limit the analysis to those configurations in which the $M$ tests
   on $X$ are performed by preparing $M$ copies of $A$ in the same initial state. 
   The QFI matrix approach~\cite{HELS67,HOLE11,ParisRev}  will then be used to evaluate
   the associated accuracy, optimizing it with respect to the initial preparation of $A$ (with respect to both its internal and external degrees of freedom) and comparing it with the results obtained in Ref.~\cite{bagan} for the case   of direct estimation. 
 Most importantly, we will also study an entanglement-assisted (EA) strategy, where each copy of $A$ 
is initialized into an entangled state of $A$ with an external ancilla system $B$, which
does not interact directly with $X$ (see Fig.~\ref{fig:schemes}), showing 
a clear improvement in the performance of the estimation process with respect to the 
non-EA (NEA) strategy in line with the findings of Refs.~\cite{quantmetr,Fujiwara}.

The paper is organized as follows. 
In Sec.~\ref{scattering potential and Cramer Rao} 
 we introduce our scattering model and  briefly review some basic aspects of  quantum estimation theory.
Then,
 in Secs.~\ref{sec:EAstrategy} and~\ref{sec:NEAstrategy} we will discuss two different strategies for the tomographic reconstruction of the target qubit, with and without the help of quantum correlations in the probe preparation. In particular, in Sec.~\ref{sec:EAstrategy}  we first derive the exact expression for
 the QFI matrix in the case of an EA configuration, in which the probe is
 initialized in a maximally entangled state with the ancilla system, and discuss some applications of the
 result in the evaluation of some functionals of the state of the target (specifically its purity and its
 azimuthal angle). In Sec.~\ref{sec:NEAstrategy}, instead, we compare the 
 EA and NEA cases by focusing on a special configuration
 in which the target state is characterized by a single unknown parameter. 
  The paper ends with Sec.~\ref{sec:conclusion} by summarizing our results in the light of future perspectives.
 The more technical aspects of the derivation are presented in a couple of appendices.


\section{The model}
\label{scattering potential and Cramer Rao}
In this section we introduce the model and set up the notation.

\subsection{1D scattering of the probe $A$}
 Suppose that a target qubit $X$ is fixed at a given position $x=0$ on a line, and that its unknown  state is described by the density matrix
\begin{equation}
\rho_X(\bm{v})=\frac{1+\bm{v}\cdot\bm{\sigma}_X}{2},\qquad 0\le|\bm{v}|\le1,
\end{equation}
with  $\bm{\sigma}=(\sigma^x,\sigma^y,\sigma^z)$ being the Pauli operators, and $\bm{v}=(v_x,v_y,v_z)$  the 3D Bloch vector, which represents the set of parameters we wish to recover via QST\@.
 In Ref.~\cite{paper:scattering} it was shown that $\bm{v}$ [and hence $\rho_X(\bm{v})$]
can be obtained 
  from the transmission and reflection probabilities of a probe qubit $A$ scattered off $X$ through 
a point-like interaction, which couples the internal degrees of freedom of the two qubits via
the following Heisenberg-type Hamiltonian
\begin{equation} \label{HH}
H=\frac{p_A^2}{2m}+g({\bm \sigma}_X\cdot{\bm \sigma}_A)\delta{(x_A)}.
\end{equation}
Here, $m$ and $p_A$ are the mass and the momentum operators of $A$, $g$ {\blue is} a positive coupling constant,  and $\bm{\s}_{J}$ are the Pauli operators acting on the qubit $J\,(=X,A)$.
  Specifically, in Ref.~\cite{paper:scattering} $A$  was assumed to be initialized into a known input state $|k\rangle\langle k| \otimes \rho_A^{\mathrm{in}}$ and  injected from the left of the line with momentum $\hbar k>0$. This kind of systems can be considered as models for a magnetic impurity spin embedded along a 1D wire \cite{ref:ImpuritySpins}, such as a semiconductor quantum wire \cite{ref:NanoWire} or a single-wall carbon nanotube \cite{ref:Tans-Nature1997}.
For instance, an electron is sent through the 1D wire as a probe particle, and its spin state is resolved after the scattering by spin-sensitive filters \cite{ref:Kawabata-SpinFilter}. Alternatively, an electron populating the lowest subband of the 1D wire can undergo scattering from two double quantum dots \cite{DQDs} to which it is electrostatically coupled.
The state tomography  of $\rho_X(\bm{v})$ then proceeded by solving the associated scattering problem and looking at  the  state of $A$ which emerges either on the left  
(transmitted component) or on the right   (reflected component), or both. 
 Such states admit a simple expression in terms of the  scattering matrix $S_{XA}$ of  the process 
 defined by  the  unitary operator
\begin{equation}
S_{XA} = \int d k \, |k \rangle \langle k| \otimes S_{XA}^{\bm{t}} 
+ \int d k\,|{-k} \rangle \langle k| \otimes S_{XA}^{\bm{r}},
\label{sigma2n}
\end{equation}
with
\begin{equation}
S_{XA}^{\bm{t}\dag} S_{XA}^{\bm{t}} +S_{XA}^{\bm{r}\dag} S_{XA}^{\bm{r}} = 1 ,
\qquad
S_{XA}^{\bm{t} \dagger} S_{XA}^{\bm{r}} +S_{XA}^{\bm{r}\dagger} S_{XA}^{\bm{t}} = 0.
\end{equation}
Here  the $|k\rangle$'s represent the momentum eigenstates of the probe $A$ ($\hbar k$ being the associated eigenvalues), while the $4\times 4$ matrices  $S_{XA}^{\bm{t}}$ and $S_{XA}^{\bm{r}}$ define the spin-dependent scattering amplitudes
associated  with  transmission and  reflection events, respectively.  They are functions of  the probe wave number $k$ and are given by 
\begin{equation}
 S_{XA}^{\bm{t},\bm{r}} =\alpha_{\bm{t},\bm{r}} (\Omega)+ \beta_{\bm{t},\bm{r}} (\Omega) (\bm{\s}_X\cdot\bm{\s}_A),
 \label{eq:Sr1}
\label{eq:SAXdef}
\end{equation}
with $\Omega$ being the dimensionless parameter 
\begin{equation}
\Omega=\frac{m g}{\hbar |k|}, 
\label{OMEGADEF}
\end{equation}
 and 
\begin{eqnarray}
 \alpha_{\bm{t}} (\Omega)=\frac{1-2i \Omega}{(1- 3i \Omega)(1+i \Omega)},\qquad \beta_{\bm{t}} (\Omega)=\frac{-i\Omega}{(1- 3i \Omega)(1+i \Omega)},
 \label{eq:St}\\
 \alpha_{\bm{r}}(\Omega)=\frac{-3 \Omega^2}{(1- 3i \Omega)(1+i \Omega)},\qquad \beta_{\bm{r}} (\Omega)=\beta_{\bm{t}} (\Omega).
 \label{eq:Sr}
\end{eqnarray}
For an explicit derivation of all these equations we refer the reader to~\cite{paper:scattering}.

Consider now an experimental  setting in which 
an observer tries to reconstruct $\rho_X(\bm{v})$ 
by merging 
 incoherently\footnote{
 By incoherent merging of the transmitted and reflected data, we mean that
no joint measurements are allowed
on the transmitted and reflected signals (a scenario which is realistic if 
the  rhs and lhs detectors are located sufficiently apart from each other).} the 
data  associated with the transmission events (to the right of the target) and the reflection events (to the left).  The  final state of $A$  can be expressed as the tensor product of   effective \emph{qutrit} density operators 
\begin{eqnarray}\label{eq:NEAtr}
\rho^{\bm{t+r}}_{A}(\bm{v}) &=&\ketbras{e^{\bm{r}}}{e^{\bm{r}}}{A} \otimes \Tr_X\{S_{XA}^{\bm{t}}[\rho_X(\bm{v})\otimes  \rho^{\mathrm{in}}_A]S_{XA}^{\bm{t}\dag}\}
 \nonumber \\&&
 {}+\Tr_X\{S_{XA}^{\bm{r}}[\rho_X(\bm{v})\otimes  \rho^{\mathrm{in}}_A]S_{XA}^{\bm{r}\dag}\} 
 \otimes \ketbras{e^{\bm{t}}}{e^{\bm{t}}}{A},
 \end{eqnarray}
where  $\Tr_X\{\cdots\}$ is the partial trace over $X$,  and
$|e^{\bm{t}}\rangle_A$ ($|e^{\bm{r}}\rangle_A$) is a vector  orthogonal to
 \emph{all} the internal spin states of $A$, which
 represents the \emph{vacuum state} associated with no
 particle reaching the rhs (lhs) detector. 
 The first term represents the contribution associated with a transmitted $A$ reaching the rhs detector (vacuum on the lhs), while the  second represents the opposite one (i.e., spin on the lhs and vacuum on the rhs). The detectors are assumed to have 100\% efficiency and no particle creation or destruction by the target is possible, so that the number of particles is conserved: every incident particle emerges \emph{either} from the left \emph{or} from the right of the target.
 
 On the other hand, if the observer collects only transmitted particles, emerging from the rhs of the target, he
 will either see nothing ($A$ being
reflected by $X$)  or see  $A$
emerging with the same momentum $\hbar k$ it had when entering the  line but with a
modified spin state due to  the interaction with $X$. 
Such configuration is described by the density operator, obtained from 
 $\rho^{\bm{t+r}}_{A}(\bm{v})$ by tracing 
out the reflected case, namely,
 \begin{eqnarray}\label{eq:NEArhot_A}
 \fl\qquad\rho^{\bm{t}}_{A}(\bm{v}) &=&  \Tr_{\bm{r}}\rho^{\bm{t+r}}_{A}(\bm{v})
 \nonumber\\
 \fl&=&\Tr_X\{S_{XA}^{\bm{t}}[\rho_X(\bm{v})\otimes  \rho^{\mathrm{in}}_A]S_{XA}^{\bm{t}\dag}\}
 + \Tr\{S_{XA}^{\bm{r}}[\rho_X(\bm{v})\otimes  \rho^{\mathrm{in}}_A]S_{XA}^{\bm{r}\dag}\} \ketbras{e^{\bm{t}}}{e^{\bm{t}}}{A}.
\end{eqnarray}
Notice that here
\begin{equation}\label{eq:NEAnonTRAsmCoeff}
\Tr\{S_{XA}^{\bm{r}}[\rho_X(\bm{v})\otimes  \rho^{\mathrm{in}}_A]S_{XA}^{\bm{r}\dag}\}
=1-\Tr\{S_{XA}^{\bm{t}}[\rho_X(\bm{v})\otimes  \rho^{\mathrm{in}}_A]S_{XA}^{\bm{t}\dag}\}
\end{equation}
is the reflection probability (i.e., the probability of no detection of $A$  on the rhs of the  line).

 An analogous expression holds for the alternative experimental setting  in which the observer only collects information of the signals emerging from the lhs of the line.  In this case Eq.~(\ref{eq:NEArhot_A})
 is replaced by  
\begin{eqnarray}\label{eq:NEArhor_A}
\fl\qquad\rho^{\bm{r}}_{A}(\bm{v}) &=&
 \Tr_{\bm{t}}\rho^{\bm{t+r}}_{A}(\bm{v})
\nonumber\\
\fl&=&
 \Tr\{S_{XA}^{\bm{t}}[\rho_X(\bm{v})\otimes  \rho^{\mathrm{in}}_A]S_{XA}^{\bm{t}\dag}\} \ketbras{e^{\bm{r}}}{e^{\bm{r}}}{A}
+ \Tr_X\{S_{XA}^{\bm{r}}[\rho_X(\bm{v})\otimes  \rho^{\mathrm{in}}_A]S_{XA}^{\bm{r}\dag}\}.
\end{eqnarray}

The output density matrices  $\rho^{\bm{t+r/t/r}}_{A}(\bm{v})$, associated with the three different
experimental settings, are functions of the state $\rho_X(\bm{v})$ of $X$. Therefore, one can acquire information on the latter by performing  QST on the former. Furthermore, since $\rho^{\bm{t+r/t/r}}_{A}(\bm{v})$  also depends on the  input state $\rho^{\mathrm{in}}_{A}$ and on the input momentum $\hbar k$ of the probe $A$, one can try to optimize the resulting accuracy with respect to these parameters. 

\subsection{Entanglement-assisted scheme}
\begin{figure}
\begin{center}
\includegraphics[height=0.33\columnwidth]{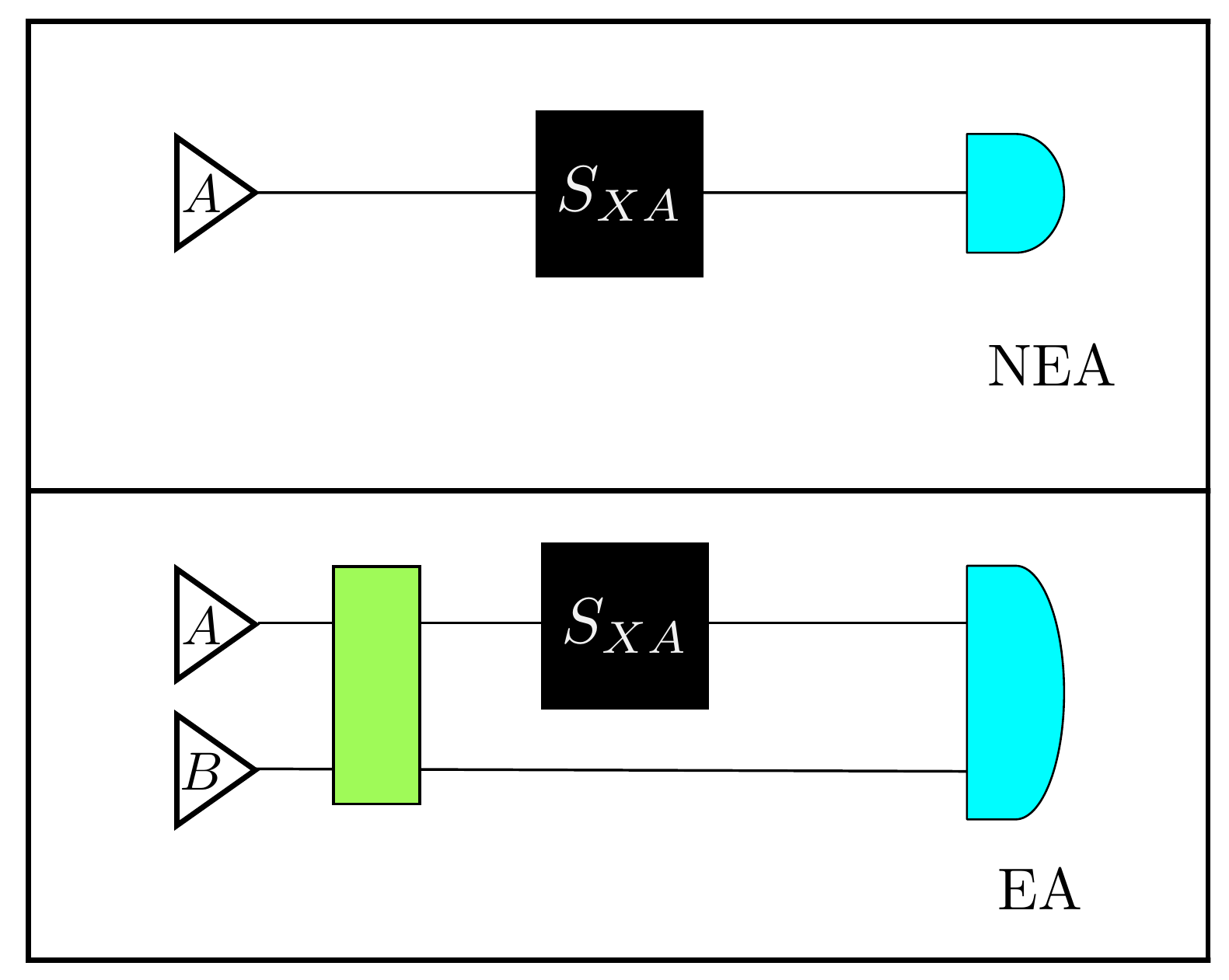}
\end{center}
\caption{(Color online) Strategies for the estimation of the target parameters. In the NEA
 strategy the probe $A$ and the ancilla system $B$ are in a separable state, and, as will be clarified in Sec. \ref{sec:NEAstrategy}, this analysis can be performed by completely neglecting $B$. On the other hand, in the 
 EA strategy  quantum correlations are introduced (rectangular box). In both schemes  one  
  makes use of optimal POVMs for the estimation of the target parameters.  Notice that only qubit $A$ undergoes a direct interaction ($S_{XA}$) with the target $X$.}\label{fig:schemes}
\end{figure}

An interesting variation of the previous schemes is obtained 
by considering  the case in which $A$ is prepared 
into  a joint (possibly entangled)  state $\rho_{AB}^{\mathrm{in}}$   with an ancilla system $B$,
 that is not directly interacting with $X$ (see  Fig.~\ref{fig:schemes}).
 Such EA configurations proved to be successful in boosting the efficiency of  several
quantum estimation~\cite{Fujiwara} and discrimination schemes~\cite{illumination}.  

Assuming that $B$ sits at rest in the laboratory
of the observer,  the resulting states of $AB$ emerging from the $AX$ interaction can again be expressed in terms of the
scattering matrix $S_{XA}$ given before.  
 Specifically Eqs.~(\ref{eq:NEAtr}), (\ref{eq:NEArhot_A}) and  (\ref{eq:NEArhor_A}) become 
\begin{eqnarray}
\label{eq:EAtr}
\fl\qquad \rho^{\bm{t+r}}_{AB}(\bm{v})&=&  \ketbras{e^{\bm{r}}}{e^{\bm{r}}}{A} \otimes 
\Tr_X\{(S_{XA}^{\bm{t}}\otimes \mathbb{I}_B)[\rho_X(\bm{v})\otimes  \rho^{\mathrm{in}}_{AB}](S_{XA}^{\bm{t}\dag}\otimes \mathbb{I}_B)\}\nonumber \\ 
\fl& &{}+ \Tr_X\{(S_{XA}^{\bm{r}}\otimes \mathbb{I}_B)[\rho_X(\bm{v})\otimes  \rho^{\mathrm{in}}_{AB}](S_{XA}^{\bm{r}\dag}\otimes \mathbb{I}_B)\}\otimes\ketbras{e^{\bm{t}}}{e^{\bm{t}}}{A},  
\label{eq:rhoABtr} \\  
\fl\qquad\rho^{\bm{t}}_{AB}(\bm{v})&=&\Tr_{\bm{r}}\rho_{AB}^{\bm{t+r}}
 \nonumber\\
 \fl &=&\Tr_X\{(S_{XA}^{\bm{t}}\otimes \mathbb{I}_B)[\rho_X(\bm{v})\otimes  \rho^{\mathrm{in}}_{AB}](S_{XA}^{\bm{t}\dag}\otimes \mathbb{I}_B) \}
 \nonumber\\
\fl &&{}+ \Tr_{XA}\{(S_{XA}^{\bm{r}}\otimes \mathbb{I}_B)[\rho_X(\bm{v})\otimes  \rho^{\mathrm{in}}_{AB}](S_{XA}^{\bm{r}\dag}\otimes \mathbb{I}_B) \} \otimes \ketbras{e^{\bm{t}}}{e^{\bm{t}}}{A},
 \label{eq:rhoABt}\\  
\fl\qquad \rho^{\bm{r}}_{AB}(\bm{v})&=&\Tr_{\bm{t}}\rho_{AB}^{\bm{t+r}}
  \nonumber\\
 \fl&=& \ketbras{e^{\bm{r}}}{e^{\bm{r}}}{A} \otimes\Tr_{XA}\{(S_{XA}^{\bm{t}}\otimes \mathbb{I}_B)[\rho_X(\bm{v})\otimes  \rho^{\mathrm{in}}_{AB}](S_{XA}^{\bm{t}\dag} \otimes \mathbb{I}_B) \}
  \nonumber\\
\fl&&{}+ \Tr_{X}\{(S_{XA}^{\bm{r}}\otimes \mathbb{I}_B)[\rho_X(\bm{v})\otimes  \rho^{\mathrm{in}}_{AB}](S_{XA}^{\bm{r}\dag}\otimes \mathbb{I}_B] \}, 
  \label{eq:rhoABr}
\end{eqnarray}
where  $\mathbb{I}_B$ stands for the identity operator on $B$. 
%

\subsection{The  Cram\'{e}r-Rao bound} 
In the following sections  we will compare the accuracy one can get 
 by reconstructing $\rho_X(\bm{v})$ through 
the EA  configurations
via   measurements  on the output states 
 $\rho_{AB}^{\bm{t+r}/\bm{t}/\bm{r}}(\bm{v})$ defined by the Eqs.~(\ref{eq:rhoABtr}), (\ref{eq:rhoABt}) and (\ref{eq:rhoABr}), with the corresponding  accuracy one achieves with 
 the NEA configurations associated with the output states
$\rho_{A}^{\bm{t+r}/\bm{t}/\bm{r}}(\bm{v})$ of Eqs.~(\ref{eq:NEAtr}), (\ref{eq:NEArhot_A}) and  (\ref{eq:NEArhor_A}). 
Differently from~\cite{paper:scattering} but in line with the approach of~\cite{bagan,ParisRev}, 
such accuracies will be evaluated by computing 
  the corresponding quantum CR bounds~\cite{HELS67,HOLE11,ParisRev}. 

We remind that   the quantum CR theorem 
establishes  a fundamental lower bound
on the  uncertainty of  any estimation strategy devised to recover the three components of~$\bm{v}$.
Specifically, assume that one has  $M$ copies of the state $\rho(\bm{v})$ which
encodes such parameters. (In our case, 
for the NEA setting $\rho(\bm{v})$ is given by the states 
$\rho_A^{\bm{t+r}/\bm{t}/\bm{r}}(\bm{v})$, depending on 
 whether the observer 
 collects 
only transmitted data, reflected data, or both. Similarly for the EA setting, where $\rho(\bm{v})$ is identified with $\rho_{AB}^{\bm{t+r}/\bm{t}/\bm{r}}(\bm{v})$).
A generic estimation strategy consists in assigning a (possibly joint) POVM measurement on the $M$ copies of the state, and a classical data processing
scheme that starting from the measurement outcome produces an estimation  $\bm{v}^\mathrm{est}=(v_1^\mathrm{est}, v_2^\mathrm{est}, v_3^\mathrm{est})$ 
 of the parameters $\bm{v}$. The uncertainty of the estimation can then be evaluated by means of the covariance  matrix
 $\Cov[\bm{v}]_{jk}=\overline{(v_j^\mathrm{est} -v_j )( v_k^\mathrm{est}-v_k)}$ ($j,k  \in \{x, y, z \}$)
obtained by averaging the distances between the real value of the parameter $\bm{v}$ and their
estimations. In this context 
the CR bound  implies that, independently from the adopted POVM and classical data processing scheme, such matrix must verify the inequality 
\begin{equation}\label{eq:boundToCov}
\Cov[\bm{v}]\ge\frac{1}{M}\bm{H}(\bm{v})^{-1}, 
\end{equation}
where $\bm{H}(\bm{v})$ is the QFI matrix of the encoding state~\cite{HELS67,ParisRev}, i.e.,
\begin{equation}\label{eq:QFIformula}
\fl\qquad  
\bm{H}_{j k}
=\sum_{n} \frac{(\partial_j \rho_n)(\partial_k \rho_n)}{\rho_n}
+2\Re\sum_{n\neq m} \frac{(\rho_n-\rho_m)^2}{\rho_n+\rho_m}
\bracket{\psi_n}{\partial_j \psi_m}\bracket{\partial_k \psi_m}{ \psi_n}.
\end{equation}
Here, $\rho(\bm{v})=\sum_{n}\rho_n\ketbras{\psi_n}{\psi_n}{}$ is the spectral decomposition of the encoding state
$\rho(\bm{v})$, while $\partial_j$ stands for the partial derivative with respect to the $j$th
component $v_j$ of the vector $\bm{v}$. 
The $j$th diagonal element of the inequality~(\ref{eq:boundToCov}) provides 
 the CR bound for the variance 
  associated with  the accuracy 
in the estimation of the  parameter $v_j$, for fixed values of the others, i.e.,\footnote{
Indeed, one can easily verify 
that the quantity $[(\bm{H}^{-1})_{jj}]^{-1}$ coincides with the QFI function  associated with the estimation of the parameter $v_j$ on the one-parameter family of states $\rho(v_j)$
obtained from $\rho(\bm{v})$  when assigning fixed values to the other components of $\bm{v}$. 
}
\begin{equation}\label{eq:boundToVar}
\Cov[\bm{v}]_{jj}=\Var[v_j]= \overline{( v_j^\mathrm{est}-  v_j )^2}\ge\frac{1}{M} (\bm{H}^{-1})_{jj}
\end{equation}
(notice  the $1/\sqrt{M}$ SQL scaling of $\sqrt{\Var[v_j]}$). 
It is worth observing that even though 
the bound~(\ref{eq:boundToCov}) is not always attainable, 
the bound~(\ref{eq:boundToVar}) is known to be asymptotically achievable for a sufficiently large $M$. 
More generally, Eq.~(\ref{eq:boundToCov}) permits  also to derive
an accuracy bound on the variance of the estimation of any 
given function $f=f(\bm{v})$  of the 
parameters $\bm{v}$~\cite{HELS67,ParisRev}. Specifically 
by re-parameterizing the problem with a  new set of independent parameters 
$\tilde{\bm{v}}= (\tilde{v}_1(\bm{v}), \tilde{v}_2(\bm{v}), \tilde{v}_3(\bm{v}))$, which
include the quantity of interest as (say) the first element $\tilde{v}_1(\bm{v}) = f(\bm{v})$, one gets, 
\begin{equation}\label{eq:boundToVarG}
\Var[f]\geq\frac{1}{M} (\tilde{\bm{H}}^{-1})_{11},
\end{equation}
where $\tilde{\bm{H}}= B \bm{H} B^T$ is the QFI matrix of the parameters $\tilde{\bm{v}}$ obtained
from $\bm{H}$ via the similarity transformation induced by the Jacobian matrix
 $B_{j k} = \partial v_k /\partial \tilde{v}_j$. As in the case of Eq.~(\ref{eq:boundToVar}), for fixed
 values of $\tilde{v}_2$ and $\tilde{v}_3$
 the inequality~(\ref{eq:boundToVarG}) establishes a bound on the accuracy reachable in the estimation
 of the function $f$ (the bound being achievable for a sufficiently large $M$ under the assumption
 that $\tilde{v}_2$ and $\tilde{v}_3$ are known \emph{a priori}).

As an application of the above construction, and for future reference, it is
instructive to report the quantum CR bounds associated with a direct 
estimation of  $\rho_X(\bm{v})$  which has been first computed in Ref.~\cite{bagan}.
In this scenario the  observer is 
assumed to have complete (not probe-mediated) access to the target state $X$, so that the encoding state $\rho(\bm{v})$
introduced above coincides with the density matrix  $\rho_X(\bm{v})$. 
In this case the QFI matrix possesses  a simple form  
 in polar coordinates,
 \begin{equation}
(v_x,v_y,v_z)=(r \sin \theta \cos \phi,r \sin \theta \sin \phi,r \cos \theta), 
\end{equation} 
where it is diagonal. Indeed using Eq.~(\ref{eq:QFIformula}) one finds,
 \begin{equation}\label{VCOVdir}
\Cov[(r, \theta, \phi)]\geq\frac{1}{M} \bm{H}^{\bm{\mathrm{dir}}}(r,\theta,\phi)^{-1},
\end{equation}
where $\bm{H}^{\bm{\mathrm{dir}}}(r,\theta,\phi)$ is the QFI matrix given by
\begin{equation}\label{VCOVdir1}
\bm{H}^{\bm{\mathrm{dir}}}(r,\theta,\phi)=\diag(c_r^{\bm{\bm{\mathrm{dir}}}},c^{\bm{\mathrm{dir}}}_\theta,c^{\bm{\mathrm{dir}}}_\theta \sin^2 \theta ),
\end{equation}
with 
\begin{equation} \label{VCOVdir2}
c_r^{\bm{\mathrm{dir}}}=\frac{1}{1-r^2}, \qquad \quad
c_\theta^{\bm{\mathrm{dir}}}=r^2.
\end{equation}
Notice that the matrix $\bm{H}^{\bm{\mathrm{dir}}}$ does not depend upon the azimuthal angle $\phi$ while it is a function  of the radial coordinate $r$ and on the polar angle $\theta$  (the  latter however  is just a geometric artifact introduced  by the polar coordinates: the north and south poles are indeed insensitive to 
rotations along the $z$-axis).

\section{Entanglement-assisted strategy}\label{sec:EAstrategy}
In this section we will consider the case in which the probe $A$ is prepared in an entangled state with the ancilla $B$, that is the EA strategy introduced in the previous section and represented in Fig.~\ref{fig:schemes}. More precisely, we will assume that subsystem $AB$ before the scattering is in a maximally entangled state, and compute the analytic expression for the associated QFI matrix $\bm{H}$ as a function of  the set of parameters defining the initial state of $X$ and of the incident momentum of $A$. In \ref{app: maximally entangled states} we will prove that our results are independent on the specific choice of the maximally entangled input state of $AB$. Henceforth we will set the input state in the singlet state:
\begin{equation}
\rho_{AB}^{\mathrm{in}}=\ketbras{\Psi^-}{\Psi^-}{AB}, \quad \kets{\Psi^-}{AB}=(\kets{0}{A}\otimes \kets{1}{B}-\kets{1}{A}\otimes \kets{0}{B})/\sqrt{2}.
\label{eq:singletdef}
\end{equation}

\subsection{Collecting data in reflection and in transmission}
Let us focus first on the experimental setting in which the observer collects both transmitted and reflected signals of the probe $A$.

From the expression~(\ref{eq:QFIformula}) of the QFI matrix it immediately follows that the reflection and transmission components of the state $\rho^{\bm{t+r}}_{AB}(\bm{v})$ defined in Eq.~(\ref{eq:EAtr})   provide two separate contributions, as they are associated to orthogonal subspaces. 
Also, as in the case of the direct estimation discussed in the previous section, it turns out that the QFI matrix possesses  a simple form  
 in polar coordinates, where it is diagonal independently of the initial state of $X$. Indeed, upon diagonalization of the
 state (\ref{eq:rhoABtr}) we find that in this case the inequality~(\ref{VCOVdir}) gets replaced by
\begin{equation} \label{VCOVEATR}
\Cov[(r, \theta, \phi)]\ge\frac{1}{M} \bm{H}^{\bm{t+r}}_{\mathrm{EA}}(r,\theta,\phi)^{-1},
\end{equation}
where   the  QFI matrix is 
\begin{equation}\label{VHTREA}
\bm{H}^{\bm{t+r}}_{\mathrm{EA}}(r,\theta,\phi)
=\diag(c^{\bm{t+r}}_r,c^{\bm{t+r}}_\theta,c^{\bm{t+r}}_\theta \sin^2\theta),
\end{equation}
with
\begin{eqnarray}
\fl \quad c^{\bm{t+r}}_r(r,\Omega)&=&\frac{8 \Omega ^2 (1+18 \Omega ^2+63 \Omega ^4)}{(1-r^2)(1+\Omega^2)(1+5 \Omega ^2)(1+9 \Omega ^2)^2}, 
\label{eq:crEAtr} \\
\fl \quad c^{\bm{t+r}}_{\theta}(r,\Omega)&=&
\frac{1}{(1+\Omega ^2)(1+9 \Omega ^2)}
\nonumber\\
\fl \quad &&
{}\times[
4(1+5 \Omega ^2)(1+9 \Omega ^2)(1+18 \Omega ^2+63 \Omega ^4)
\nonumber\\
\fl\quad&&\qquad\qquad\qquad\quad
{}-r^2(1+4 \Omega ^2+68 \Omega ^4+720 \Omega ^6+1863 \Omega ^8)  
]\nonumber \\
 \fl \quad &&\times
\frac{32 r^2 \Omega ^2}{ [4(1+9 \Omega ^2)^2-r^2(1-9 \Omega ^2)^2][4 (1+5 \Omega^2)^2-r^2(1+3\Omega
   ^2)^2]}
   \label{eq:cthetaEAtr}
\end{eqnarray}
(in~\ref{app: QFIcartesian} we also report the expression of the bound in cartesian coordinates). 
As in the case of the direct estimation, 
 $\bm{H}^{\bm{t+r}}_{\mathrm{EA}}$ does not depend upon the azimuthal angle $\phi$, while it is a function  of the radial coordinate $r$ and the polar angle $\theta$. Such a behavior is associated with the symmetry of the coupling Hamiltonian~(\ref{HH}), which does not possess a preferred spatial direction,
 and of the input
state of $AB$.\footnote{We stress that the dependence of $\bm{H}^{\bm{t+r}}_{\mathrm{EA}}$ on $\theta$ has nothing to do with the probe-target coupling: as in the case of 
$\bm{H}^{\bm{\mathrm{dir}}}$ it is a geometric artifact 
 of the polar
representation.}
It is also worth pointing out that  the matrix $\bm{H}^{\bm{t + r}}_{\mathrm{EA}}$ vanishes 
 when $\Omega$ is zero or infinite  (that is, infinite or zero incident momentum $\hbar k$ of the probe $A$). This implies that  for such configurations
 no recovering of information on $X$ is  possible.
 Indeed if $k=0$ it means
 that $A$ is initially at rest and will not be able to be scattered by $X$, while for very large $k$ it means that $A$ propagates so fast along the  line that 
 the interaction with $X$ can only have a minimal  
 (asymptotically vanishing) impact on its evolution.

\begin{figure}
\begin{center}
\includegraphics[width=0.5\columnwidth]{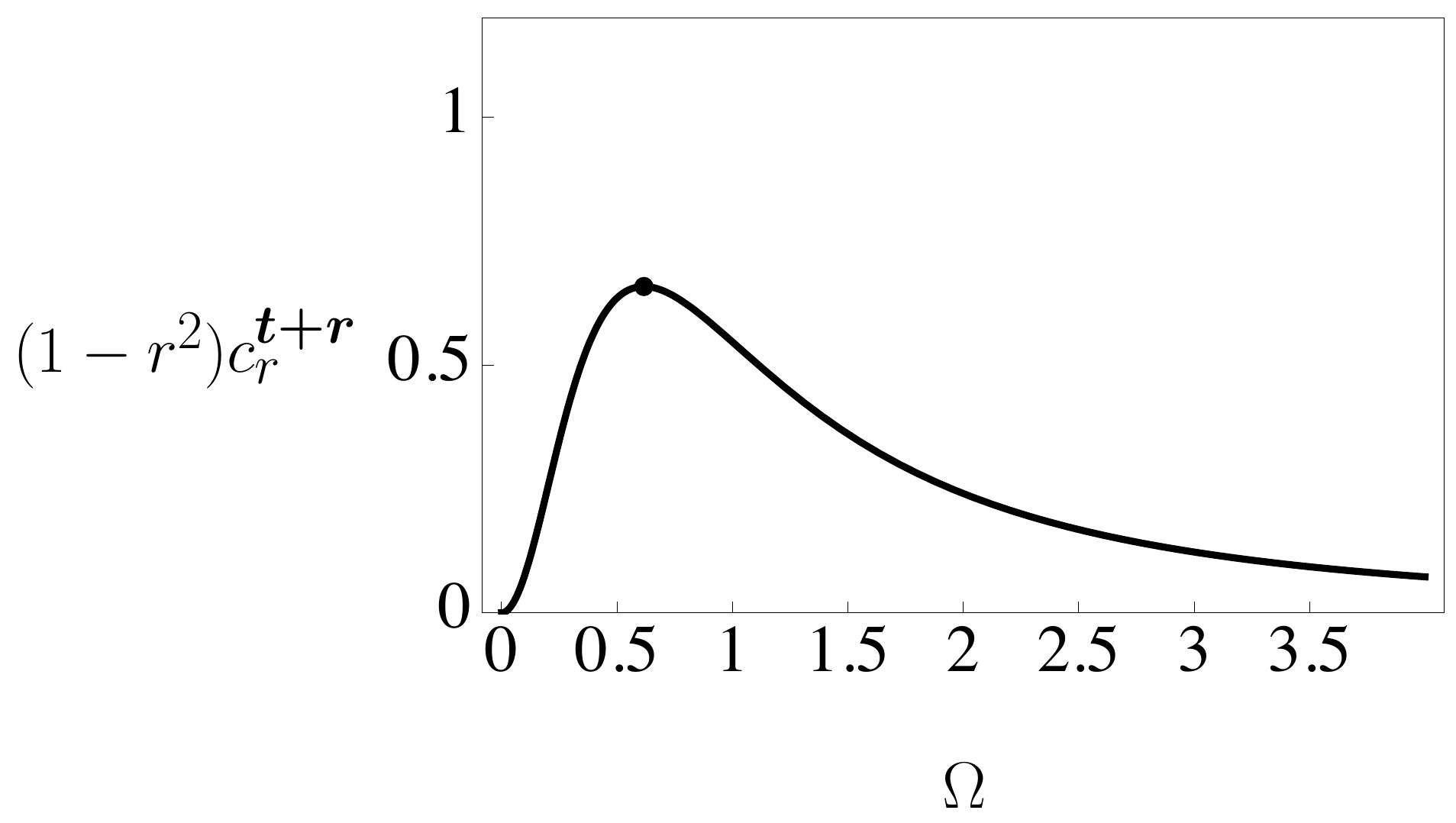}
\caption{Functional dependence of the rescaled  QFI function $(1-r^2)c^{\bm{t+r}}_r$, whose inverse  bounds  the accuracy~(\ref{PURIAE}) achievable in the determination of $r$ 
with  an EA strategy where $AB$ is initialized in a maximally entangled state and data are collected both in transmission and in reflection cases. 
The dot indicates the optimal incoming momentum for  probe $A$.}\label{fig:HQQTRASMREF}
\end{center}
\end{figure}

As a  specific example suppose then that the observer, already knowing the value of the parameters $
\theta$ and $\phi$, is interested in recovering the missing  
parameter $r$, which determines the purity of the target system $X$, $\Tr\rho^2_X(\bm{v}) = r^2$. 
Equation~(\ref{VCOVEATR}) then yields
\begin{equation} \label{PURIAE}
\Var[r]\ge\frac{1}{M}\frac{1}{c^{\bm{t+r}}_r(r,\Omega)} = \frac{1-r^2}{M}
 \frac{(1+\Omega ^2)(1+5 \Omega ^2)(1+9 \Omega ^2)^2}{8 \Omega ^2(1+18 \Omega ^2+63 \Omega ^4)}
,
\end{equation}
which should be compared with the quantity $(1/M) (1/c^{\bm{\mathrm{dir}}}_r)= (1-r^2)/M$ one obtains in the
direct estimation case, i.e., Eqs.~(\ref{VCOVdir})--(\ref{VCOVdir2}). 
Since the rational function of $\Omega$ on the rhs of Eq.~(\ref{PURIAE}) is larger than $1$, it follows
that $(1/M ) (1/c^{\bm{\mathrm{dir}}}_r)$  is always smaller than the EA bound (this is very much expected  since in the case of direct estimation
the observer has access to the $X$ system, while in the EA strategy he can only  recover info on $X$ through 
the probe $A$).  The minimum 
of the rhs Eq.~(\ref{PURIAE})  is reached when $\Omega\simeq0.616$ as shown in Fig.~\ref{fig:HQQTRASMREF}, this value provides the optimal input momentum $\hbar k$ of  $A$ via Eq.~(\ref{OMEGADEF}), for  recovering the parameter~$r$. For such a choice the EA strategy misses the direct
estimation accuracy just by  a  factor of~$1.52$. 
Notice finally that for both the EA strategy and for the direct one, the accuracy
bounds vanish with  the $\sqrt{1-r^2}$ distance from the  surface of the Bloch sphere (this is a consequence of the fact that it is intrinsically simpler to  distinguish pure states from 
 mixed states). 

Consider next the case in which the observer, already knowing the value of the parameters $r$ and $\theta$, 
is interested in recovering the   azimuthal phase $\phi$  of the target system $X$. 
This is given by the third diagonal element of the QFI matrix~(\ref{VHTREA}),~i.e.,
\begin{equation} \label{phiAE}
\Var[\phi]\geq\frac{1}{M} \frac{1}{c^{\bm{t+r}}_\theta(r,\Omega) \sin^2\theta} ,
\end{equation}
which again should be compared with the bound 
$(1/M)(1/c^{\bm{\mathrm{dir}}}_\theta \sin^2 \theta)= (1/M)(1/r^2 \sin^2 \theta)$ one gets for the direct estimation. Again one can verify that the rhs of Eq.~(\ref{phiAE}) is always larger than the 
direct estimation threshold (notice however that both expressions diverge for $\theta=0$ and $\theta=\pi$ due to fact that for such choices the variable $\phi$ is not even  defined).
For the sake of simplicity we focus on the special case in which $X$ is in a pure state  on the equatorial plane of the Bloch sphere (i.e., $\theta=\pi/2$ and $r=1$). 
In this case Eq.~(\ref{phiAE}) yields
\begin{equation} \label{phiAE1}
\Var[\phi]\geq\frac{1}{M} 
   \frac{3 (1+\Omega ^2)(1+3 \Omega ^2)(1+7 \Omega^2)(1+9 \Omega ^2)}{32  \Omega^2(1+10 \Omega ^2+27 \Omega^4)}, 
\end{equation}
which reaches its minimum when $\Omega\simeq 0.637$, where the rhs is $\simeq 1.354/M$, falling short  of the direct threshold $1/M$ by  35\%.


\begin{figure}
\begin{center}
\includegraphics[width=1\columnwidth]{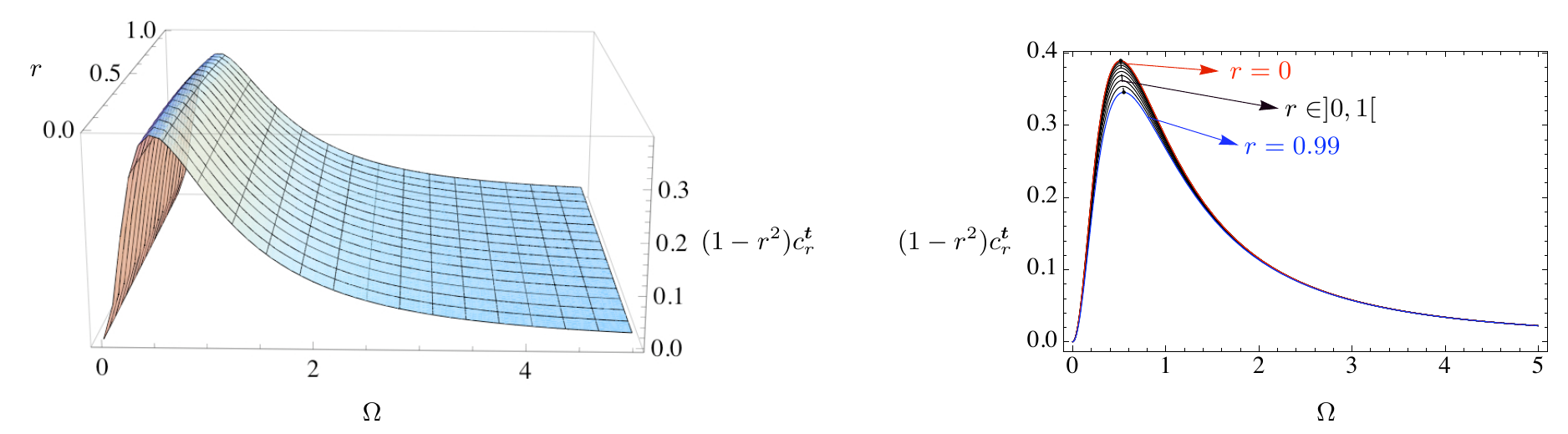}
\caption{(Color online) Plots of the rescaled QFI coefficient $(1-r^2) {c^{\bm{t}}_r(r,\Omega)}$ of Eq.~(\ref{eq:crEAt}) for the transmission case as a function of $r$ and $\Omega$. For every value of $r$ there exists an optimal value of the incident momentum $k$. The upper and lower curves in the right panel refer to the case of  completely mixed (i.e., $r=0$) and almost pure states (i.e., $r\simeq 1$), respectively.}\label{fig:crTRASM}
\end{center}
\end{figure}
\subsection{Collecting only reflected or transmitted data} 
By repeating the same analysis for the case in which only transmitted or reflected probes are detected (see Eqs.~(\ref{eq:rhoABt}) and (\ref{eq:rhoABr}))
 we find that Eq.~(\ref{VCOVEATR}) still holds with 
the QFI  matrix $\bm{H}^{\bm{t+r}}_{\mathrm{EA}}(r,\theta,\phi)$ being replaced by
\begin{eqnarray}
\bm{H}^{\bm{t}/\bm{r}}_{\mathrm{EA}}(r,\theta,\phi)=\diag({c}^{\bm{t}/\bm{r}}_r,c^{\bm{t}/\bm{r}}_\theta,c^{\bm{t}/\bm{r}}_\theta \sin^2\theta),
\end{eqnarray}
with
\begin{eqnarray}
\fl \quad {c}^{\bm{t}}_r(r,\Omega)&=&\frac{2 \Omega ^2[3(1+3 \Omega ^2)(11+76 \Omega ^2+117 \Omega ^4)-2 r^2(9+50 \Omega ^2+45 \Omega ^4)]}{(1-r^2)(1+\Omega^2)(1+5 \Omega ^2)(1+9 \Omega^2)[9(1+3 \Omega ^2)^2-4 r^2]},\label{eq:crEAt}\\
\fl \quad c^{\bm{t}}_\theta(r,\Omega)&=&\frac{4 r^2 \Omega ^2[2(1+5 \Omega ^2)(11+76 \Omega ^2+117 \Omega ^4)-r^2(1+3  \Omega ^2)^2]}{3(1+\Omega^2)(1+3 \Omega^2)(1+9 \Omega^2)[4(1+5 \Omega^2)^2-r^2(1+3 \Omega^2)^2]}, \label{eq:cthetaEAt}
 \end{eqnarray}
in transmission, while
\begin{eqnarray}
\fl \quad c^{\bm{r}}_r(r,\Omega)&=&\frac{2 \Omega ^2 [(1+7\Omega^2)(1+36\Omega^2+207\Omega^4)-2r^2\Omega^2(1+18\Omega^2+117\Omega^4)]}{(1-r^2)(1+\Omega^2)
(1+9 \Omega^2)^2[(1+7\Omega^2)^2-4r^2\Omega ^4]},\label{eq:crEAr}\\
\fl \quad c^{\bm{r}}_\theta(r,\Omega)&=&\frac{4 r^2 \Omega^2[2(1+9\Omega^2)(1+36\Omega^2+207\Omega ^4)-r^2\Omega ^2(1-9\Omega^2)^2]}{(1+\Omega^2)(1+7\Omega^2)(1+9\Omega^2)[4(1+9 \Omega^2)^2-r^2(1-9 \Omega ^2)^2]},\label{eq:cthetaEAr}
 \end{eqnarray} 
in reflection.

\begin{figure}
\begin{center}
\includegraphics[width=1\columnwidth]{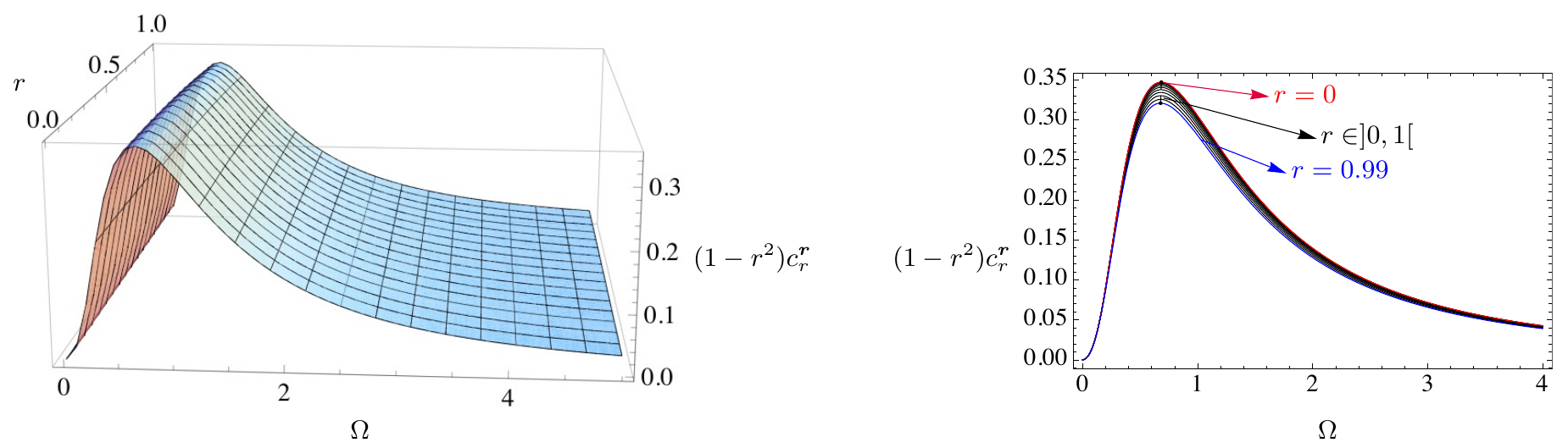}
\caption{(Color online) As in 
Fig.~\ref{fig:crTRASM}, when only reflected data are collected.}\label{fig:crREFL}
\end{center}
\end{figure}

\begin{figure}
\begin{center}
\includegraphics[width=0.6\columnwidth]{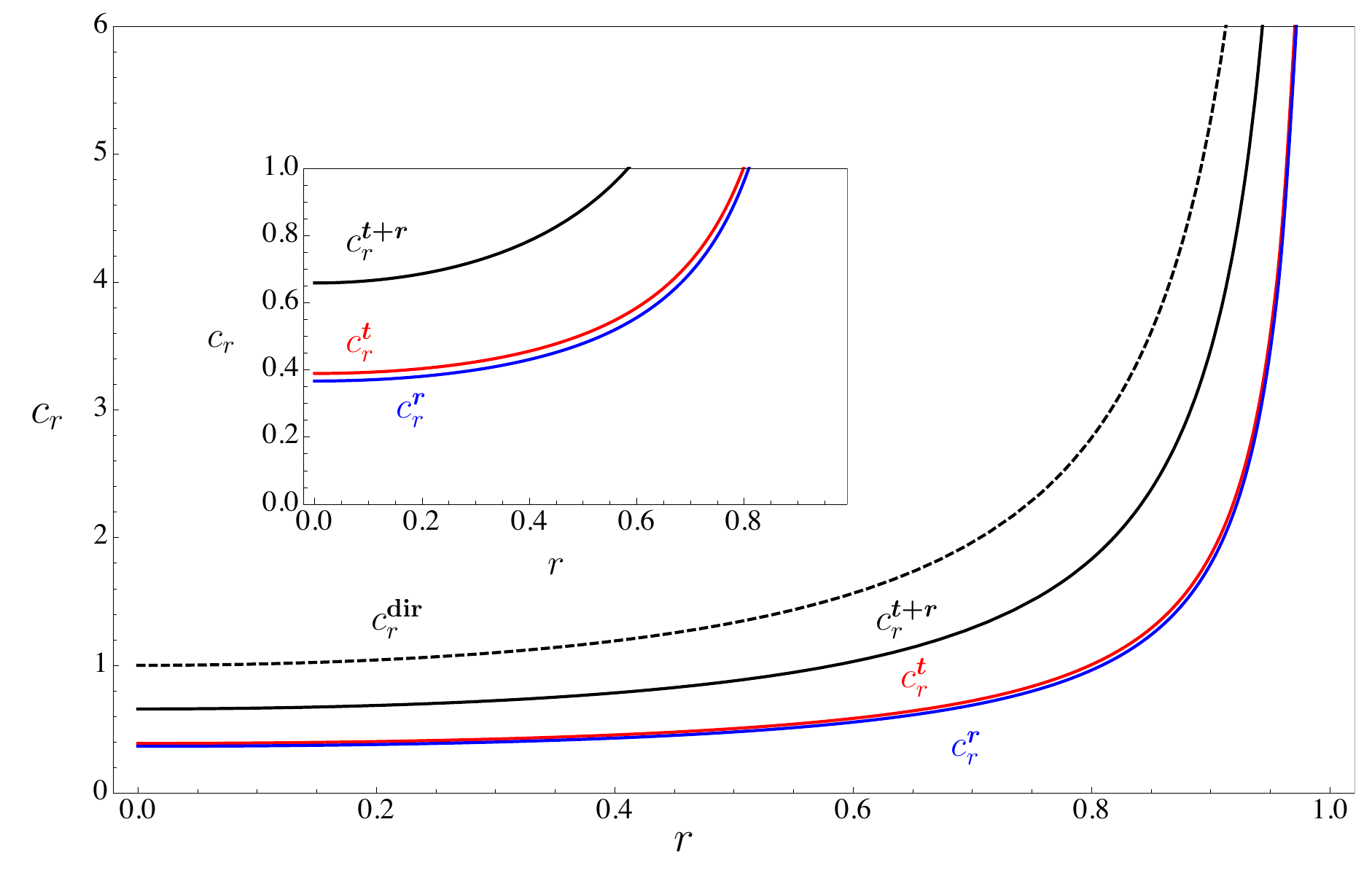}
\end{center}
\caption{(Color online) Plot of the quantum CR bound for rescaled variance
$M  \Var[r]$, of  the estimation of the  radius $r$ of the Bloch sphere describing the target qubit, in polar coordinates. The dashed line refers to the ultimate limit achievable  by direct access to the target, while the
solid lines refer to the case in which both reflected and transmitted, only transmitted and only reflected data are collected.
In particular,  the case of both 
reflected and transmitted data is plotted at the optimal 
value $\Omega\simeq 0.61$, which, in this case, is independent of the value of $r$. The case in which only reflected or only transmitted data are collected has been plotted  instead by considering, for each given $r$, 
the corresponding optimal value of $\Omega$. 
Notice that for all values of $r$, we find $c_r^{\bm{t+r}}>c_r^{\bm{t/r}}$, since by collecting all the scattering data we gain the largest amount  of information on the target. Futhermore, observe that  transmission and  reflection  give almost the same result, with a little improvement of transmission  with respect to  reflection.}\label{fig:crQ}
\end{figure}

Notice that, in this case, even if there is still a quadratic divergence of $c^{\bm{r}}_r$ and $ c^{\bm{t}}_r$ at $r=1$,  their dependence on $r$ does not factorize. Hence
in the estimation of the radius $r$, the bound (\ref{PURIAE}) is replaced by a lower bound, $(1/M)(1/c^{\bm{t}/\bm{r}}_r)$,
which  for every value of $r$ admits an optimal momentum for the incident probe $A$ that maximizes the associated QFI, as shown in Figs.~\ref{fig:crTRASM}  and~\ref{fig:crREFL}. 
Interestingly enough, however,  there exist ``optimality intervals" for the incident momentum (i.e., $\Omega \in [0.51,0.55]$ for  transmission, and  $\Omega \in [0.67,0.68]$ for  reflection), which
guarantee rather high performances for all values of $r$. 
Observe that in order to achieve the optimal  estimation from  transmitted data we have to send the probe faster than  the case of reflection (recall that $\Omega \propto 1/k$): this is in perfect agreement with the intuitive idea that to efficiently collect the data in transmission
$A$ should be given a sufficiently large initial momentum to prevent back scattering from $X$
(and vice versa for reflection). Between the above two intervals lies the optimal value of $\Omega$ for the case in which we collect all  scattering data, $\Omega \simeq  0.61$.\\ 
In Fig. \ref{fig:crQ} we plot the QFI of the radius $r$ (purity) for the case of transmitted, reflected, or both reflected and transmitted data. Of course, the last case leads to the best estimation of the target purity, as we are collecting the largest amount of information from the scattering process. 

Analogous considerations apply also to the estimation of the azimuthal angle $\phi$.
One can easily extract them by a direct comparison of the coefficients (\ref{eq:cthetaEAt}) and (\ref{eq:cthetaEAr}) with the corresponding expression associated with the transmission \emph{and} reflection strategy given in Eq.~(\ref{eq:cthetaEAtr}) and also with the direct estimation procedure [see Eq.~(\ref{VCOVdir2})].

\section{Comparing EA and NEA strategies}\label{sec:NEAstrategy}
In this section we will present a comparison between the EA strategies introduced in the previous 
section with the NEA strategies obtained by restricting the  analysis to the case in which the initial state $\rho_{AB}$  of the probe $A$ and the ancilla $B$ is separable. 
For the sake of simplicity, we find it instructive to restrict the analysis to the 
situation in which,  as in Fig.~\ref{fig:crQ}, the observer aims only to 
 estimate the purity of the target state $X$. 
 Specifically we will work under the assumption that the three dimensional   vector $\bm{v}$
 which specifies $\rho_X(\bm{v})$ in the Bloch sphere  lies on the
 $z$ axis, that is  
 $v_x=v_y=0$ and  
 \begin{equation}
\rho_X(v_z)=\frac{1+v_z  {\sigma}_X^{z}}{2},\qquad v_z\in[-1,1].
\end{equation}
Under this condition the quantum CR bound~(\ref{eq:boundToCov})
reduces to an inequality for the variance $\Var[v_z]$ of 
the unique parameter $v_z$, i.e.,
\begin{equation} \label{PURIAEz}
\Var[v_z]\ge\frac{1}{M}H_\mathrm{EA/NEA}^{\bm{t+r}/\bm{t}/\bm{r}}(v_z)^{-1}
,
\end{equation}
where  $H_\mathrm{EA/NEA}^{\bm{t+r}/\bm{t}/\bm{r}}(v_z)$ are the QFI  functions of the problem 
computed as usual by exploiting the spectral decomposition of the output states $AB/A$ and
 using Eq.~(\ref{eq:QFIformula}) (which in this case contains only derivatives with respect
 of the unique parameter $v_z$). 
 
For the EA configuration the functions  $H_{\mathrm{EA}}^{\bm{t+r}/\bm{t}/\bm{r}}(v_z)$  
 coincide with the third diagonal elements of the QFI matrices in  cartesian coordinates  evaluated at $v_x=v_y=0$, namely,
 \begin{eqnarray}
 \fl \qquad H_{\mathrm{EA}}^{\bm{t+r}}(v_z)&=&
 \frac{8\Omega^2(1+18\Omega^2+63\Omega^4)}{(1-v_z^2)(1+\Omega^2)(1+5\Omega^2)(1+9 \Omega^2)^2}, 
 \\ 
 \fl \qquad H_{\mathrm{EA}}^{\bm{t}}(v_z)&= &
 \frac{2 \Omega^2[3(1+3\Omega^2)(11+76\Omega^2+117\Omega^4)-2v_z^2(9+50\Omega^2+45\Omega^4)]}{(1-v_z^2)(1+\Omega^2)(1+5\Omega^2)(1+9 \Omega^2)[9(1+3\Omega^2)^2-4 v_z^2]} , 
   \\ 
\fl \qquad H_{\mathrm{EA}}^{\bm{r}}(v_z)&=&\frac{2\Omega^2[(1+7\Omega^2)(1+36\Omega^2+207\Omega^4)-2v_z^2\Omega^2(1+18\Omega^2+117\Omega^4)]}{(1-v_z^2)(1+\Omega^2)(1+9\Omega^2)^2[(1+7\Omega^2)^2-4v_z^2\Omega^4]}.  
 \end{eqnarray}
 Notice that the above expressions can be retrieved from Eqs. (\ref{eq:crEAtr}), (\ref{eq:crEAt}) and (\ref{eq:crEAr}) by substituting $r$ with $v_z$. 
The general expression of the QFI matrix for the EA configuration can be found in~\ref{app: QFIcartesian}\@.

 Before computing the QFI functions for the NEA configuration,  it is worth observing that 
due to the convexity of the QFI function (e.g., see Ref.~\cite{Fujiwara}), 
 it follows that  in the NEA  scenario the contribution of the ancilla $B$ in the scattering 
 can be completely neglected, see 
 Fig.~\ref{fig:schemes}. Indeed, since the maximum of a QFI function is  always achieved 
 on pure states, one can restrict the analysis of the NEA configuration to  pure separable 
 states of $AB$: for them however the output states of $AB$ also factorize and 
  one  can completely ignore the subsystem $B$.
Consequently,  the states of the system after the scattering 
are given by  Eqs.~(\ref{eq:NEAtr}), (\ref{eq:NEArhot_A}) or (\ref{eq:NEArhor_A}), 
for the case in which we collect either all scattering data, or only transmitted/reflected probes. 
In particular we will assume
 \begin{equation}
\rho_A^{\mathrm{in}}=\frac{1+\bm{n}\cdot \bm{\sigma}_A}{2}, \qquad \bm{n}=(\sin \theta_A , 0,\cos \theta_A),
\end{equation}
where, without loss of generality, the input azimuthal angle $\phi_A$ has been set to zero by using the
 symmetry of the coupling Hamiltonian.

Let us then consider first the accuracy achievable when collecting data only on transmission, i.e.,
assuming as output state the one given in Eq.~(\ref{eq:NEArhot_A}). 
The resulting QFI is the following (involved) function of $\theta_A$:
\begin{eqnarray}
\fl H^{\bm t}_{\mathrm{NEA}}(v_z,\theta_A,\Omega) 
&=&
[
4(11+96\Omega^2+181\Omega^4)
-v_z^2(1+\Omega ^2)(1+33 \Omega ^2)
   \nonumber\\
   \fl
   &&\qquad
   {}   
   -4 v_z(8+43 \Omega ^2+3 \Omega ^4)\cos\theta_A
-4(1-\Omega^2+8v_z^2\Omega^2)(1+\Omega^2)\cos2\theta_A
\nonumber\\
   \fl
   &&\qquad\qquad
   {}
   -4v_z\Omega^2(1+\Omega^2)\cos3\theta_A
   +v_z^2(1+\Omega ^2)^2\cos4\theta_A
   ]
   \nonumber\\ 
   \fl \qquad\qquad &&\times\frac{1}{4(1+5\Omega^2)-v_z^2(1+17\Omega^2)-v_z(1+\Omega^2)(4\cos\theta_A-v_z\cos2\theta_A)}
   \nonumber\\
   \fl&& \times\frac{\Omega ^2}{(1+\Omega^2)(3+9\Omega^2-2v_z\cos\theta_A)(1+7\Omega^2+2v_z\Omega^2\cos\theta_A)}.
   \end{eqnarray}
 Still one recognizes some general features we already observed  in the previous section. 
In particular the QFI function vanishes at $\Omega=0$ and $\Omega =\infty$.  
Notice also that    
when the target state is pure, there exists a particular choice of the initial state of the probe such that the above quantity diverges quadratically, as found for the  EA strategy, namely  $v_z=1, \theta_A=0$ and $v_z=-1, \theta_A=\pi$, and arbitrary (finite) $\Omega>0$. 
Finally, notice that for each value of $v_z$ there exists an optimal point for the pair $(\theta_A,\Omega)$  that maximizes the QFI\@. 
\begin{figure}
\begin{center}
\includegraphics[height=0.28\columnwidth]{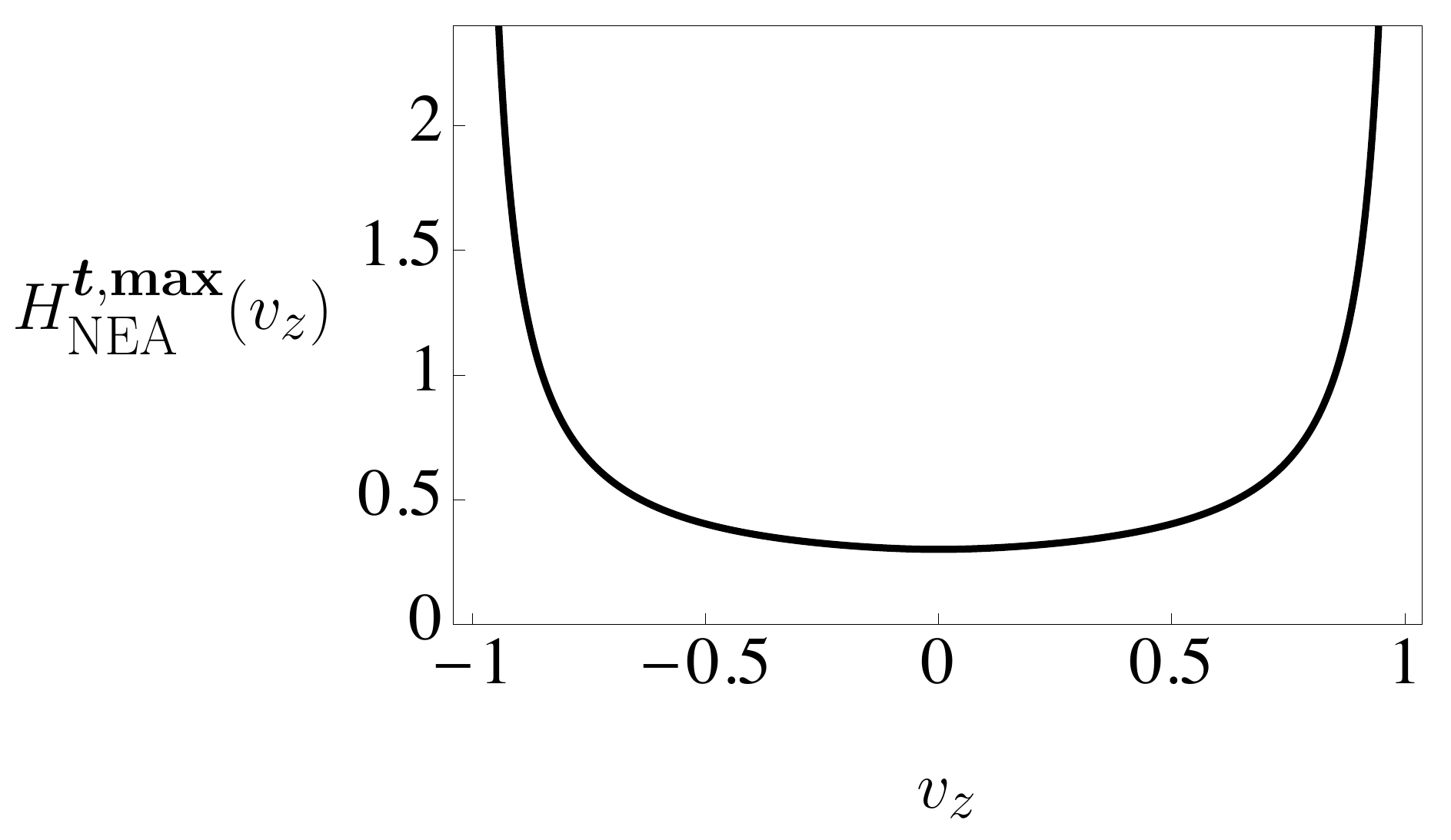}
\caption{Envelope of the QFI for the NEA configuration corresponding to the optimal choice of $\theta_A$ and $\Omega$ for each value of $v_z$, if only transmitted data are collected. Notice the divergence at $v_z=\pm1$, which for all $\Omega$ corresponds to $\theta_A=0$ and $\pi$.}
\label{fig:HcqBest}
\end{center}
\end{figure}
In Fig.~\ref{fig:HcqBest} we plot the envelope of the QFI for $v_z \in [-1,1]$, that is the maximum value of $H^{\bm t}_{\mathrm{NEA}}(v_z)$, which can be proved to be a symmetric function of $v_z$. Notice that the initial state of $A$  that yields the best estimation of the target state  depends on $v_z$, that is, on the initial state of the target, which is in principle unknown. We can only say that once we have set the direction of the probe $A$ in the Bloch sphere before the scattering, the best optimization we can get involves pure target states with Bloch vectors parallel to this direction. The same analysis can be repeated for  reflection and yields:
\begin{eqnarray}
\fl&&H^{\bm r}_{\mathrm{NEA}}(v_z,\theta_A,\Omega)
\nonumber\\
\fl &&
\quad
=[
4(5+23\Omega^2)
-v_z^2(1+\Omega ^2)-4v_z(3-2\Omega ^2)\cos\theta_A
\nonumber\\
\fl &&\quad\qquad\quad
{}+4(1-5\Omega ^2)\cos2\theta_A
-4v_z(1+2\Omega^2)\cos3\theta_A
+v_z^2(1+\Omega ^2)\cos4\theta_A
]
   \nonumber\\
   \fl&&
   \quad\quad
   {}\times \frac{1}{3(1+3\Omega^2)(1+7\Omega^2)-2v_z^2\Omega^2-2v_z(1+4 \Omega ^2-9 \Omega ^4)\cos\theta_A-2v_z^2\Omega ^2\cos2\theta_A}
\nonumber\\ 
   \fl&&
      \quad\quad
\times\frac{\Omega^2}{(1+\Omega^2)(4-v_z^2-4v_z\cos\theta_A+v_z^2\cos2\theta_A)}.
\end{eqnarray}
Similarly to the case of transmission  the optimal values of  $\theta_A$ and $\Omega$ are symmetric functions of $v_z$. Moreover, when the target state is pure, i.e., $v_z=  1$ and $-1$, the QFI diverges for $\theta_A=0$ and $\pi$, respectively, as for  transmission.
\begin{figure}
\begin{center} \qquad \qquad \quad
\includegraphics[height=0.3\columnwidth]{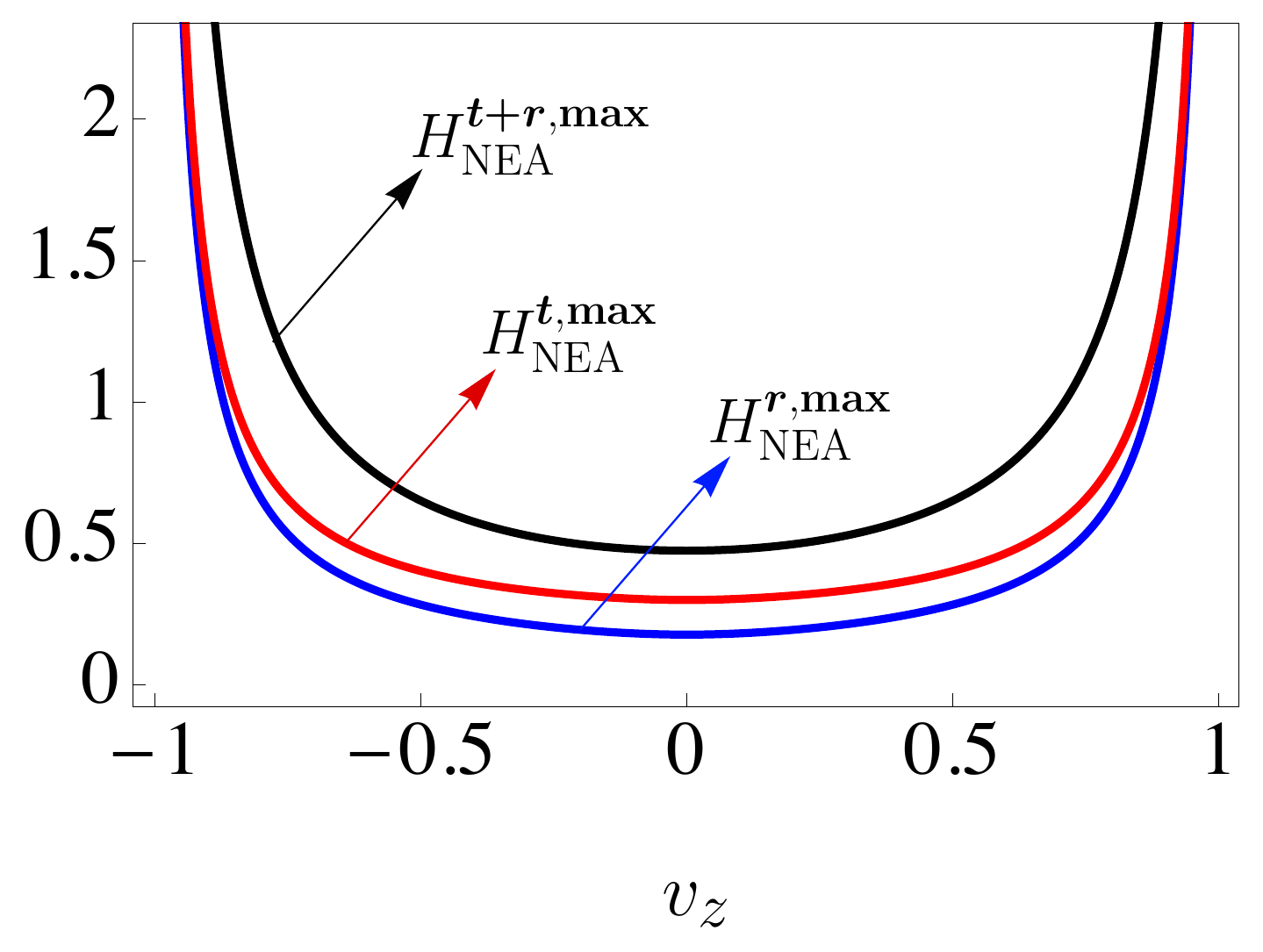}
\caption{(Color online) Envelope of the QFI for the NEA strategy corresponding to the optimal choice of $\theta_A$ and $\Omega$ for each value of $v_z$, for the case in which we collect either transmitted, or reflected or all scattered probes. Notice that the QFI diverges for $v_z=1$ and $-1$ when $\theta_A=0$ and $\pi$, respectively.} \label{fig:HCQ}
\end{center}
\end{figure}

Finally, the largest amount of information on the system can be inferred by collecting both transmitted and reflected data. In this case the QFI associated to the state (\ref{eq:NEAtr}) of the incident probe after the scattering is given by
\begin{eqnarray}
\fl & & H^{\bm {t+r}}_{\mathrm{NEA}}(v_z,\theta_A,\Omega)
\nonumber\\
\fl&&
\quad
=\{
(1+7\Omega^2+2v_z\Omega^2\cos\theta_A)
   \nonumber\\ 
\fl   &&\qquad\quad
{}\times
[4(1+5\Omega^2)-v_z^2(1+17\Omega^2)
-v_z(1+\Omega^2)(4\cos\theta_A-v_z \cos2\theta_A)]
   \nonumber\\ 
   \fl&& \qquad \quad
   \times[4(5+27\Omega^2)-v_z^2+4(1-9\Omega^2)\cos2\theta_A-16v_z\cos{\theta_A}^3 +v_z^2\cos4\theta_A]
   \nonumber\\ 
\fl   &&\qquad\quad
+(4-v_z^2-4v_z\cos\theta_A+v_z^2\cos2\theta_A)
[3(1+3\Omega^2)-2v_z\cos\theta_A]
   \nonumber\\ 
\fl   &&\qquad\qquad
   \times[
   4(3+48\Omega^2+181\Omega^4)
   -v_z^2\Omega^2(1+33\Omega^2)
   -12v_z\Omega^2(1+\Omega ^2)\cos\theta_A
   \nonumber\\ 
\fl   &&\qquad\qquad\quad
   -4(1+8\Omega^2-\Omega^4+8 v_z^2\Omega^4)\cos2\theta_A
   -v_z\Omega^2(1+\Omega ^2)(4\cos3\theta_A-v_z \cos4\theta_A)]
   \}
   \nonumber\\
\fl && \quad\qquad
{}\times\frac{1}{4(1+5\Omega^2)-v_z^2(1+17\Omega^2)-v_z(1+\Omega^2)(4 \cos\theta_A-v_z \cos2 \theta_A)}
   \nonumber\\ 
\fl && \quad\qquad
{}\times\frac{1}{
   (4-v_z^2-4v_z\cos\theta_A+v_z^2\cos2\theta_A)(3+9 \Omega^2-2v_z\cos\theta_A)}
   \nonumber\\
\fl && \quad\qquad
   \times\frac{\Omega ^2}{(1+\Omega ^2)(1+9\Omega^2)(1+7\Omega^2+2v_z\Omega^2\cos\theta_A)}.
\end{eqnarray}
This function exhibits the same properties as $H^{\bm t}_{\mathrm{NEA}}$ and $H^{\bm r}_{\mathrm{NEA}}$. The three curves plotted in Fig.~\ref{fig:HCQ}, all symmetric with respect to $v_z$, refer to the optimal values of the QFI for the three cases analyzed above. As expected, the collection of all  scattering data returns the best tomographic reconstruction of the target state.  Notice that the transmission strategy seems to overcome the reflection one, like in the EA strategy. Also it can be shown  that for all $v_z$ the optimal value of $\Omega$ for the transmission is lower than that for reflection, and for the case in which we collect both  transmitted and reflected data it is in between (this results have been obtained by numerical optimizations).

\begin{center}
\begin{figure}
\begin{center}
\includegraphics[width=0.9\columnwidth]{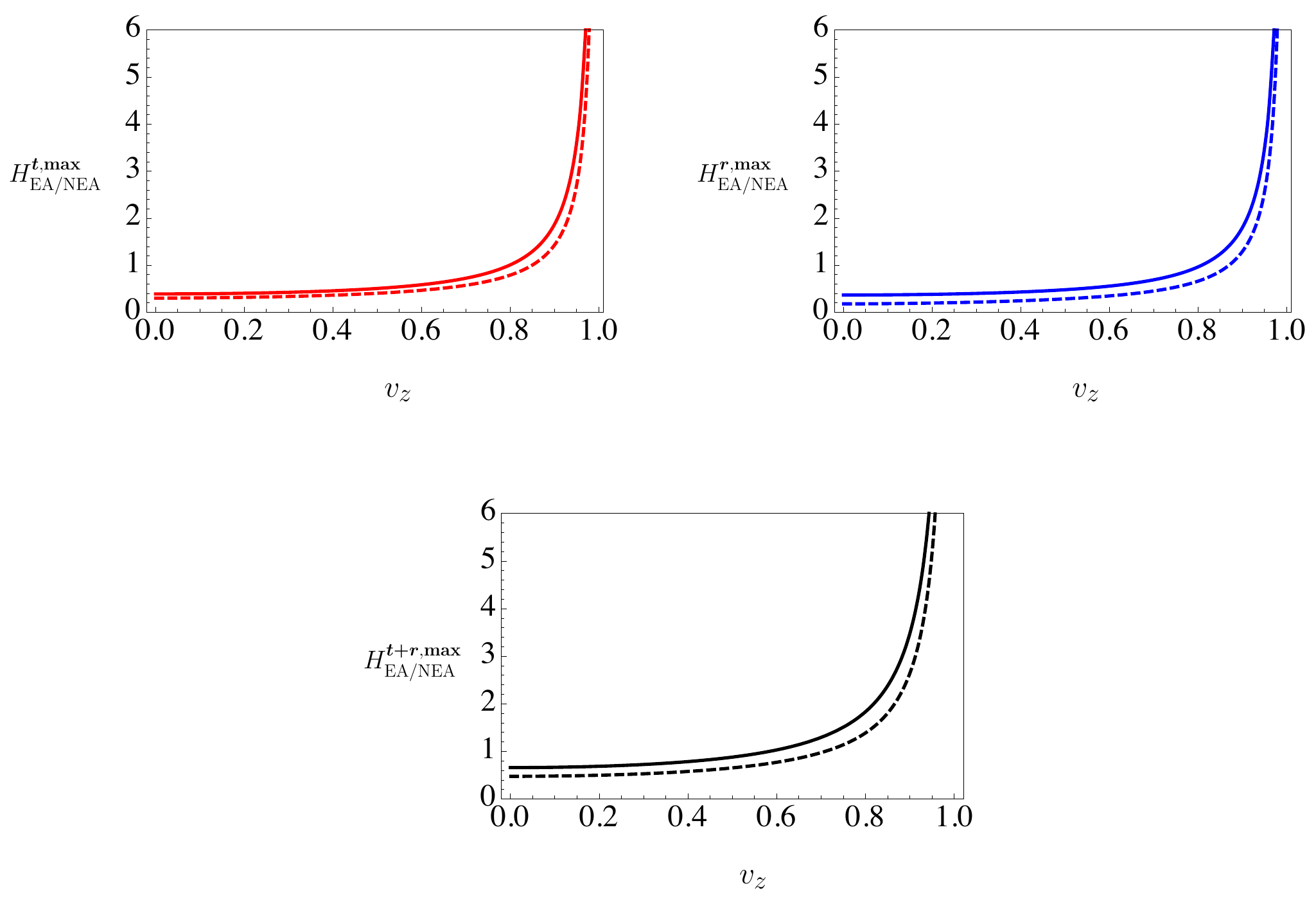}
\caption{(Color online) Maximum value of the QFI associated to the parameter $v_z$ of the target qubit $X$. The solid line refers to the EA strategy, where the probe $A$ is in a maximally entangled state with the ancilla $B$, while the dashed line refers to the NEA.}\label{fig:EA_NEA}
\end{center}
\end{figure}
\end{center}

In order to compare the efficiency of the NEA strategies with the EA strategies discussed in the previous section, in Fig.~\ref{fig:EA_NEA} we  plot the maximum QFI for $v_z \geq 0$, for the EA  (solid line) and the NEA (dashed line) strategies (the expression for the EA strategies have been obtained by exploiting the representation of the QFI matrix in cartesian coordinates reported in~\ref{app: QFIcartesian}). The role played by the entanglement between the probe and the ancilla before the scattering is evident: it implies an enhancement in the QFI for all $v_z$, both for the case in which we collect all the scattering data and for the case in which we have access only to trasmission/reflection events. 
 \section{Conclusion}\label{sec:conclusion}

We have presented a detailed study of the tomographic state reconstruction of a target system, obtained by monitoring a scattered probe. 
Focusing on the special case in which both the target and the probe are qubit systems, and assuming 
the scattering  to take place on a 1-D line, we used quantum estimation techniques to evaluate the efficiency of the process 
in several configurations of interest. In particular we distinguished two regimes:
the EA regime in which the observer is allowed to initialize the probe in an entangled state with an external ancilla that
it is kept in the laboratory; and the NEA regime where instead no entanglement is allowed between the probe and the ancilla.
As expected, when all the other settings are kept identical, the EA strategies turn out to be more effective then their NEA counterparts.

Within both regimes we have also studied what happens when the observers have access to all or only part of the scattered data. Specifically we consider
the cases in which only transmitted or reflected data are used in the tomographic reconstruction, noticing that these regimes are characterized by different
optimal values for the input momentum of the probing particle (the transmitted scenario being characterized by higher optimal  input momenta than the reflected one). 
We have also analyzed the situation in which both transmitted and reflected data are available to the observer, assuming though that no joint coherent
measurements could be performed on the associated quantum degree of freedom (i.e. we explicitly excluded the possibility of performing joint detection on the
left and right side of the 1-D channel, an hypothesis which is very much reasonable if the detectors are located at sufficient large distances from each other). 
The overall accuracy clearly benefits from this possibility: still it remains below the threshold~\cite{bagan} one gets when direct access to the target system is allowed.
An open problem is to the determine whether or not one could exploit other sort of quantum resources to close such gap. A possible candidate which we aim to explore in a future
development of the work, is to allow 
the observer to \emph{entangle} different probes together (in the present scenario indeed, even though we used $M$ probes, they were all prepared in a factorized configuration 
of the {\em same} initial state). This strategy could in fact  benefit from super-additivity effects arising from the presence of non classical correlations between the various probes, resulting
in higher performances of the tomographic reconstruction. 
An interesting open problem is also to understand to what extend the specific features we have observed in our simple scattering model (qubits interacting along a 1-D line via
Heisenberg-like coupling) could be generalize to more complex configurations (say by increasing the size the of the target and/or of the probe, or by  allowing the scattering to take place in a 2-D or a 3-D
setting).

This work was supported by the Italian Ministry of University and Research 
through FIRB-IDEAS Project No.\ RBID08B3FM
as well as
under the bilateral Italian-Japanese Projects II04C1AF4E on ``Quantum Information, Computation and Communication,''
by the Joint Italian-Japanese Laboratory on ``Quantum Information and Computation'' of the Italian Ministry for Foreign Affairs,
and
by a Special Coordination Fund for Promoting Science and Technology and a Grant-in-Aid for Young Scientists (B) (No.\ 21740294) both from the Ministry of Education, Culture, Sports, Science and Technology, Japan.

\appendix
\section{Equivalence between maximally entangled states}\label {app: maximally entangled states}
We explicitly prove that our analysis for the EA strategies is independent of the particular choice of a maximally entangled state for the subsystem $AB$. 

The Heisenberg-type coupling $\bm{\sigma}_X \cdot \bm{\sigma}_A$  describing the interaction between qubits $X$ and $A$ can be written in terms of the so-called \textit{swap operator} $\mathcal{S}_{X|A}$ as
\begin{equation}
\bm{\sigma}_X \cdot \bm{\sigma}_A = 2\mathcal{S}_{X|A} -  \mathbb{I}_{XA}, \qquad \mathcal{S}_{X|A}=\sum_{i,j}  \ketbras{i}{j}{X} \otimes \ketbras{j}{i}{A}.
\end{equation}
It is a unitary self-adjoint operator and is characterized by the following very simple property
\begin{equation}
(\mathbb{I}_X\otimes U_A)\mathcal{S}_{X|A}(\mathbb{I}_X\otimes U_A^\dag) 
= (U_X^\dag\otimes \mathbb{I}_A)\mathcal{S}_{X|A}'(U_X\otimes \mathbb{I}_A),
\end{equation}
where $U_X$ and $U_A$ act in the same way on $\mathcal{H}_X$ and $\mathcal{H}_A$, respectively ($U_X=U_A$), and $\mathcal{S}_{X|A}'$ is the swap operator in the rotated frame
\begin{equation}
\mathcal{S}_{X|A}'=\sum_{i,j}  \ketbras{r_i}{r_j}{X} \otimes \ketbras{r_j}{r_i}{A}, \qquad \kets{r_i}{X/A} = U_{X/A} \kets{i}{X/A}.
\end{equation}
The above property is trivially conserved for the interaction Hamiltonian $\bm{\sigma}_X \cdot \bm{\sigma}_A$.
The equivalence among all the maximally entangled input states for the EA strategies is thus straightforward. 
Indeed,  a generic maximally entangled state of the probe $A$ and the ancilla $B$ can always be written as $\kets{\Psi_{UV}}{AB}=(U_A^\dag\otimes V_B^\dag)\kets{\Psi^-}{AB}$, with $\kets{\Psi^-}{AB}$
the singlet state~(\ref{eq:singletdef}). If we send $|\Psi_{UV}\rangle_{AB}$ to the target qubit $X$ it can be easily shown that the contributions given by $\mathrm{Tr}_X\{\ldots\}$ in the final state of the subsystem $AB$ become
\begin{eqnarray}\label{eq:rho_ab_fin max}
\nonumber\fl &&(U_A^\dag \otimes V_B^\dag)\mathrm{Tr}_X[(U_A S^{t/r}_{XA}{U_A}^{\dagger})(\rho_X(\bm{v})\otimes|\Psi^-\rangle_{AB}\langle\Psi^-|)({U_A}{S^{t/r}_{XA}}^\dag U_A^{\dagger})](U_A\otimes
         V_B) \nonumber\\
\nonumber\fl &&=(U_A^\dag \otimes V_B^\dag)\mathrm{Tr}_X[{S'^{t/r}}_{XA}(\rho'_X(\bm{v})\otimes|\Psi^-\rangle_{AB}\langle\Psi^-|)S^{\prime{t/r}\dag}_{XA}](U_A\otimes V_B) \end{eqnarray} with
\begin{eqnarray}
S_{XA}^{\prime\bm{t},\bm{r}} =\alpha_{\bm{t},\bm{r}} (\Omega)+ \beta_{\bm{t},\bm{r}} (\Omega) (2\mathcal{S'}_{X|A} - \mathbb{I}_{XA}) ,\quad \rho'_X= U_X  \rho_X U_X^{\dagger}. \nonumber \end{eqnarray}
Since the physical properties of the system do not depend on the choice of the reference basis, the results of our analysis for the EA strategies are completely independent of the choice of an initial maximally entangled state of $AB$.

\section{QFI matrix in cartesian coordinates}\label{app: QFIcartesian}
 In this appendix 
 we consider the explicit expression of the Fisher matrix for the EA strategy and discuss the symmetry properties of its elements.

For the case in which both transmitted and reflected probes are collected, the off diagonal elements of the Fisher matrix are given by
 \begin{eqnarray}
\fl
&&[\bm{H}^{\bm{t+r}}_{\mathrm{EA}}(v_x,v_y,v_z)]_{ij,i \neq j}
\nonumber\\
\fl&&\quad
=\frac{2 v_i v_j\Omega^2}{(1- \bm{v}^2)(1+\Omega^2)(1+5\Omega^2)(1+9\Omega^2)^2}\nonumber\\
\fl&&\quad\qquad
\times
\{(1+7\Omega^2)(1+9\Omega^2)(3+13\Omega^2)^2[
4(1+9\Omega^2)^2
-\bm{v}^2(1-9\Omega^2)^2
]\nonumber\\
\fl&&\quad\qquad\qquad
+3(1+3\Omega^2)(1+5\Omega^2)(1+27\Omega^2)^2
[
4(1+5\Omega^2)^2
-\bm{v}^2(1+3\Omega^2)^2]
\}\nonumber\\
\fl&&\quad\qquad
\times
\frac{1}{
[4(1+9\Omega^2)^2
-\bm{v}^2(1-9\Omega^2)^2
]
[4(1+5\Omega^2)^2-\bm{v}^2(1+3\Omega^2)^2]
},
 \end{eqnarray}
with $i,j \in \{x,y,z\}$,
 while for the diagonal elements we have
\begin{eqnarray}
\fl&&
[\bm{H}^{\bm{t+r}}_{\mathrm{EA}}(v_x,v_y,v_z)]_{ii}
\nonumber\\
\fl&&\quad
= \frac{2\Omega^2}{
  (1-\bm{v}^2)(1+\Omega^2)(1+5\Omega^2)(1+9\Omega^2)^2
  } 
   \nonumber
   \\ 
  \fl &&\quad\qquad
  \times
 \{
(1+9\Omega^2)(3+13\Omega^2)
[
4(1+9\Omega^2)^2
-\bm{v}^2(1-9\Omega^2)^2
]\nonumber\\
\fl&&\quad\qquad\qquad\qquad
\times
[
v_i^2(1+7\Omega^2)(3+13\Omega^2)
+4(1-\bm{v}^2)(1+5\Omega^2)^2
]\nonumber\\
   \fl&&\quad\qquad\qquad
   +(1+5\Omega^2)(1+27\Omega^2)
   [
   4(1+5\Omega^2)^2-\bm{v}^2(1+3\Omega^2)^2
   ]
   \nonumber\\
\fl&&\quad\qquad\qquad\qquad
   \times
   [
  3v_i^2(1+3\Omega^2)(1+27\Omega^2)
  +4(1-\bm{v}^2)(1+9\Omega^2)^2
   ]
   \}
   \nonumber\\
   \fl&&\quad\qquad
   \times\frac{1}{
   [4(1+9\Omega^2)^2-\bm{v}^2(1-9\Omega^2)^2]
   [4(1+5\Omega^2)^2-\bm{v}^2(1+3\Omega^2)^2]
   },
 \end{eqnarray}
 with $i,j \in \{x,y,z\}$.
Notice that if there is only one out  of the three parameters characterizing the initial state of the target different from zero, $v_i \neq0$ and $v_j=v_k=0$, the QFI matrix is diagonal and thus its $ii$ element coincides with the single parameter QFI for $v_i$. Furthermore, due to the symmetry of the Heinsenberg-type interaction and of the singlet state of $A$ and $B$ before the scattering, we find
\begin{equation}\label{eq:isotropy1}
(1-{v_i}^2)[\bm{H}^{\bm{t+r}}_{\mathrm{EA}}(v_x,v_y,v_z)]_{ii}=(1-r^2)c_r^{\bm{t+r}}, \qquad i \in \{x,y,z\}.
\end{equation}
If we are able to detect only transmitted or reflected probes we get 
\begin{eqnarray}
\fl&&
[\bm{H}^{\bm{t}}_{\mathrm{EA}}(v_x,v_y,v_z)]_{ij,i \neq j}
\nonumber\\
\fl&&
\quad
=
\frac{2v_iv_j\Omega^2}{3(1-\bm{v}^2)(1+\Omega^2)(1+3\Omega^2)(1+5\Omega^2)(1+9\Omega^2)} \nonumber\\
\fl&&\quad\qquad
\times\{
[3(1+3\Omega^2)(1+7\Omega^2)(3+13\Omega^2)^2[9(1+3\Omega^2)^2-4\bm{v}^2]\nonumber\\
\fl&&\quad\qquad\qquad
+8(1-\bm{v}^2)(1+5\Omega^2)[4(1+5\Omega^2)^2-\bm{v}^2(1+3\Omega^2)^2]\} \nonumber\\
\fl&&\quad\qquad
\times
\frac{1}{[9(1+3\Omega^2)^2-4\bm{v}^2]
   [4(1+5\Omega^2)^2-\bm{v}^2(1+3\Omega^2)^2]},
\\ \nonumber\\ \nonumber\\
\fl&&
[\bm{H}^{\bm{t}}_{\mathrm{EA}}(v_x,v_y,v_z)]_{ii}
\nonumber\\
\fl&&\quad
=\frac{2\Omega^2}{3(1-\bm{v}^2)(1+\Omega^2)(1+3\Omega^2)(1+5\Omega^2)(1+9\Omega^2)}\nonumber\\
\fl&&\quad\qquad
\times\{
2(1-\bm{v}^2)(1+5\Omega^2)[4(1+5\Omega^2)^2-\bm{v}^2(1+3\Omega^2)^2]\nonumber\\
\fl&&\qquad\quad\quad\qquad
\times[9(1+3\Omega^2)^2-4(v_j^2+v_k^2)]\nonumber\\
\fl&&\quad\qquad\qquad
+3(1+3\Omega^2)(3+13\Omega^2)[9(1+3\Omega^2)^2-4\bm{v}^2]\nonumber\\
\fl&&\quad\qquad\qquad\qquad
\times[4(1+5\Omega^2)^2(1-v_j^2-v_k^2)-(1+3\Omega^2)^2v_i^2]
\}\nonumber\\
\fl&&\quad\qquad
\times \frac{1}{[9 (1+3\Omega^2)^2-4 \bm{v}^2]
   [4(1+5\Omega^2)^2-\bm{v}^2(1+3\Omega^2)^2]},
\end{eqnarray}
and
\begin{eqnarray}
\fl&&
[\bm{H}^{\bm{r}}_{\mathrm{EA}}(v_x,v_y,v_z)]_{ij,i \neq j}
\nonumber\\
\fl&&\quad
= 
\frac{2v_i v_j\Omega^2}{(1-\bm{v}^2)(1+\Omega^2)(1+7\Omega^2)(1+9\Omega^2)^2} 
\nonumber  \\
\fl &&\quad\qquad
{}\times
\{
3(1+3\Omega^2)(1+7\Omega^2)(1+27\Omega^2)^2 
[(1+7\Omega^2)^2-4\bm{v}^2\Omega^4] 
\nonumber  \\
\fl &&\quad\qquad\qquad
-8(1-\bm{v}^2)\Omega ^6(1+9\Omega^2)[4(1+9\Omega^2)^2-\bm{v}^2(1-9\Omega^2)^2]
\} 
\nonumber  \\
\fl &&\quad\qquad
\times\frac{1}{[
(1+7\Omega^2)^2-4\bm{v}^2\Omega^4] 
[4(1+9\Omega^2)^2-\bm{v}^2(1-9\Omega^2)^2]},
   \\
\fl&&
[\bm{H}^{\bm{r}}_{\mathrm{EA}}(v_x,v_y,v_z)]_{ii}
\nonumber\\
 \fl  &&\quad
 =
 \frac{2\Omega^2}{(1-\bm{v}^2)(1+\Omega^2)(1+7\Omega^2)(1+9\Omega^2)^2} 
\nonumber \\
\fl&&\quad\qquad
\times\{
(1+7\Omega^2)(1+27\Omega^2)
[(1+7\Omega^2)^2-4\bm{v}^2\Omega^4]
\nonumber \\ 
\fl &&\quad\qquad\qquad\qquad
{}\times
[4(1+9\Omega^2)^2(1-v_j^2-v_k^2)-v_i^2(1-9\Omega^2)^2]
\nonumber\\ 
\fl &&\quad\qquad\qquad
+2(1-\bm{v}^2)\Omega^2(1+9\Omega^2)[4(1+9\Omega^2)^2-\bm{v}^2(1-9\Omega^2)^2] \nonumber\\ 
\fl &&\qquad\qquad\qquad\quad \times
 ((1+7\Omega^2)^2-4\Omega^4( v_j^2-v_k^2))\}
   \nonumber\\ 
\fl &&\quad\qquad
\times\frac{1}{[(1+7\Omega^2)^2-4\bm{v}^2\Omega^4]
[4(1+9\Omega^2)^2-\bm{v}^2(1-9\Omega^2)^2]}, 
\end{eqnarray}
$\forall i,j,k \in \{x,y,z\}$. 
Analogously to the case in which we detect all the scattered probes, we find that also for the transmission and the reflection cases if we have $v_i \neq 0$ and $v_j=v_k=0$, $i\neq j \neq k \in \{x,y,z\}$, the QFI for $v_i$ is diagonal, and furthermore an equation similar to (\ref{eq:isotropy1}) holds:
\begin{equation}\label{eq:isotropy2}
(1-{v_i}^2)[\bm{H}^{\bm{t/r}}_{\mathrm{EA}}(v_x,v_y,v_z)]_{ii}=(1-r^2) c_r^{\bm{t/r}}, \qquad i \in \{x,y,z\}.
\end{equation}

\section*{}

\end{document}